\documentclass{emulateapj}
\usepackage{lscape}

\begin{document}

\shortauthors{Luhman}
\shorttitle{Stellar Population in Chamaeleon~I}

\title{The Stellar Population of the Chamaeleon~I Star-Forming 
Region\altaffilmark{1}}

\author{K. L. Luhman}
\affil{Department of Astronomy and Astrophysics,
The Pennsylvania State University, University Park, PA; kluhman@astro.psu.edu.}

\altaffiltext{1}{Based on observations performed with the Magellan
Telescopes at Las Campanas Observatory, Cerro Tololo Inter-American Observatory 
(CTIO), Gemini Observatory, and the NASA/ESA {\it Hubble Space Telescope}.
CTIO is operated by the
Association of Universities for Research in Astronomy (AURA) under a
contract with the National Science Foundation (NSF). Gemini Observatory 
is operated by AURA under a
cooperative agreement with the NSF on behalf of the Gemini partnership: the
NSF (United States), the Particle Physics and Astronomy
Research Council (United Kingdom), the National Research Council (Canada),
CONICYT (Chile), the Australian Research Council (Australia), CNPq (Brazil)
and CONICET (Argentina).
The {\it Hubble} observations are associated with proposal ID 10138 and
were obtained at the Space Telescope Science Institute, which is operated by
AURA under NASA contract NAS 5-26555. 
}

\begin{abstract}
I present a new census of the stellar population in the 
Chamaeleon~I star-forming region.
Using optical and near-IR photometry and followup spectroscopy, I have 
discovered 50 new members of Chamaeleon~I, 
expanding the census of known members to 226 objects. 
Fourteen of these new members have spectral types later than M6, which
doubles the number of known members that are likely to be substellar. 
I have estimated extinctions, luminosities, and effective temperatures for
the known members, used these data to construct an H-R diagram for the cluster,
and inferred individual masses and ages with the theoretical evolutionary 
models of Baraffe and Chabrier. 
The distribution of isochronal ages indicates that star formation
began 3-4 and 5-6~Myr ago in the southern and northern subclusters, 
respectively, and has continued to the present time at a declining rate. 
The IMF in Chamaeleon~I reaches a maximum at a mass of 0.1-0.15~$M_\odot$, 
and thus closely resembles the IMFs in IC~348 and the Orion Nebula Cluster. 
In logarithmic units where the Salpeter slope is 1.35, 
the IMF is roughly flat in the substellar regime and shows no indication of 
reaching a minimum down to a completeness limit of 0.01~$M_\odot$. 
The low-mass stars are more widely distributed than members at other masses
in the northern subcluster, but this is not the case in the southern
subcluster. Meanwhile, the brown dwarfs have the same spatial distribution
as the stars out to a radius of $3\arcdeg$ (8.5~pc) from the center
of Chamaeleon~I.
\end{abstract}

\keywords{infrared: stars --- stars: evolution --- stars: formation --- stars:
low-mass, brown dwarfs --- stars: luminosity function, mass function ---
stars: pre-main sequence}

\section{Introduction}
\label{sec:intro}

The characteristics of the distributions of masses, ages, and positions 
in a newborn stellar population are determined by the process of star 
formation. As a result, measurements of these distributions in star-forming 
regions are potentially valuable for testing models of the birth of stars and 
brown dwarfs. For instance, the properties of the stellar initial mass function 
\citep[IMF,][]{mey00}
can constrain the relative importance of turbulent fragmentation
\citep{pn02}, gravitational fragmentation \citep{lar85}, 
dynamical interactions \citep{bon03}, and accretion and outflows \citep{af96}
in regulating the final masses of stars.
Models of the star formation rates of molecular clouds (e.g., constant,
accelerating, bursts) can be tested against the distributions of 
ages and positions of members of young clusters \citep{fei96,pal97,har01}.
The spatial distributions also provide
insight into cloud fragmentation, binary formation, cluster dynamics, and
the origin of brown dwarfs \citep{lar95,hil98,har02,luh06tau}. 
To obtain measurements of this kind, one must identify the
members of star-forming regions and estimate their masses and ages. 
Only a few young stellar populations have been characterized in detail,
such as the Orion Nebula Cluster \citep{hil97}, Taurus \citep{kh95},
and IC~348 in Perseus \citep{luh03ic}.

The Chamaeleon~I star-forming region is amenable to a thorough census of 
its stellar population for several reasons. It is among the nearest 
star-forming regions \citep[$d=160$-170~pc,][]{whi97,wic98,ber99}, 
exhibits less extinction than many young clusters ($A_V\lesssim5$), 
is compact enough that a large fraction of the region can be surveyed to great 
depth in a reasonable amount of time, and is sufficiently rich for a
statistically significant analysis of its stellar population.
In addition, because Chamaeleon~I is relatively isolated from other 
star-forming regions, confusion with other young populations is minimal. 
Previous surveys have already identified more than 150 young
stars and brown dwarfs in Chamaeleon~I 
through photometric variability, H$\alpha$ emission, X-ray emission, 
mid-infrared (IR) excess emission, and optical and near-IR color-magnitude 
diagrams \citep[][references therein]{com04,luh04cha,luh07cha}.
However, in the census of known members produced by these surveys, the
completeness as a function of mass and position is unknown \citep{luh04cha}. 
In this paper, I present a set of magnitude-limited surveys for members of 
Chamaeleon~I that have well-defined completeness limits
(\S~\ref{sec:new}). I then use the new census of Chamaeleon~I to 
measure the star formation history (\S~\ref{sec:hr}), IMF (\S~\ref{sec:imf}), 
and spatial distribution of its stellar population (\S~\ref{sec:spatial}).

\section{New Members of Chamaeleon I}
\label{sec:new}

\subsection{Selection of Candidate Members}
\label{sec:cand}

I have used optical and near-IR broad-band photometry from several sources
to perform four surveys for new members of Chamaeleon~I. The fields 
covered by these surveys are indicated in the maps of the cluster in 
Figure~\ref{fig:map1}. The first survey (hereafter named IK1) is
shallow ($I\sim18$) and covers most of the cloud complex, concentrating
on the central $1.5\arcdeg\times0.35\arcdeg$ area where most of the known
members reside.  The second survey (RI) also focuses on this central field,
but with greater depth ($I\sim21$). These two surveys are designed to 
produce an IMF measurement that has good number
statistics and completeness to reasonably low masses ($M\sim0.02$~$M_\odot$). 
A third survey (IK2) considers a small range of masses 
($M=0.2$-0.02~$M_\odot$) across a very wide field extending well beyond the 
Chamaeleon~I cloud. This survey will be used to measure the spatial 
distribution of low-mass stars and brown dwarfs. Finally, a very deep survey
is performed toward a small field in the center of the southern subcluster 
in Chamaeleon~I (IZ), which will be used to constrain the minimum mass 
of the IMF. The candidate members identified in this section and observed 
spectroscopically in \S~\ref{sec:spec} are listed in Table~\ref{tab:phot},
which includes all of the relevant photometric measurements.

\subsubsection{IK1 Survey}
\label{sec:ik1}

For the IK1 survey, I used measurements in $J$, $H$, and $K_s$ from the 
Point Source Catalog of the Two-Micron All-Sky Survey \citep[2MASS,][]{skr06}
and $i$-band data from the Second Release of the Deep Near-Infrared Survey of 
the Southern Sky \citep[DENIS,][]{ep99} for the area covered by the
map of Chamaeleon~I in Figure~\ref{fig:map1}a. DENIS data are not available 
within two strips that are indicated in the map.
Using these data, I constructed an extinction-corrected diagram of $H$ versus
$i-K_s$ in the manner previously done for Chamaeleon~I by \citet{luh04cha}. 
I defined a boundary that followed the lower envelope of the sequence of
previously known members and identified objects above this boundary 
as candidate members.  I have performed spectroscopy on 112 of the $\sim500$
candidates (\S~\ref{sec:spec}) and classified them as either members
or nonmembers (\S~\ref{sec:class}). These spectroscopic targets
and the previously known members appearing in these
data are shown in the color-magnitude diagram in Figure~\ref{fig:ik1}.
When selecting candidates for spectroscopy, 
I gave highest priority to ones that are within the central
$1.5\arcdeg\times0.35\arcdeg$ area indicated in Figure~\ref{fig:map1}a 
because I will measure an IMF from members within this field (\S~\ref{sec:imf}).
The boundaries of this field were designed to encompass a significant fraction
of the known cluster members, which favors a larger area, while 
minimizing contamination by field stars among the candidates, which
favors a smaller area. 

Based on the spectral classifications performed in \S~\ref{sec:class},
the 112 IK1 candidates that were observed spectroscopically 
consist of 32 members and 80 nonmembers. 
Eight of these members were independently discovered by \citet{com04}.
In this work, these objects are counted as previously known members rather 
than new members. The primary in the wide binary
brown dwarf 2MASS~J11011926-7732383 (hereafter 2M~1101-7732) is also one of
the 32 members found during the IK1 survey, but for the purposes
of this study, I count it as a previously known member because 
I presented its discovery in a separate study \citep{luh04bin}. 
Given these considerations, the IK1 survey produced 23 new members.

\subsubsection{IK2 Survey}
\label{sec:ik2}

\begin{figure}
\epsscale{1.2}
\plotone{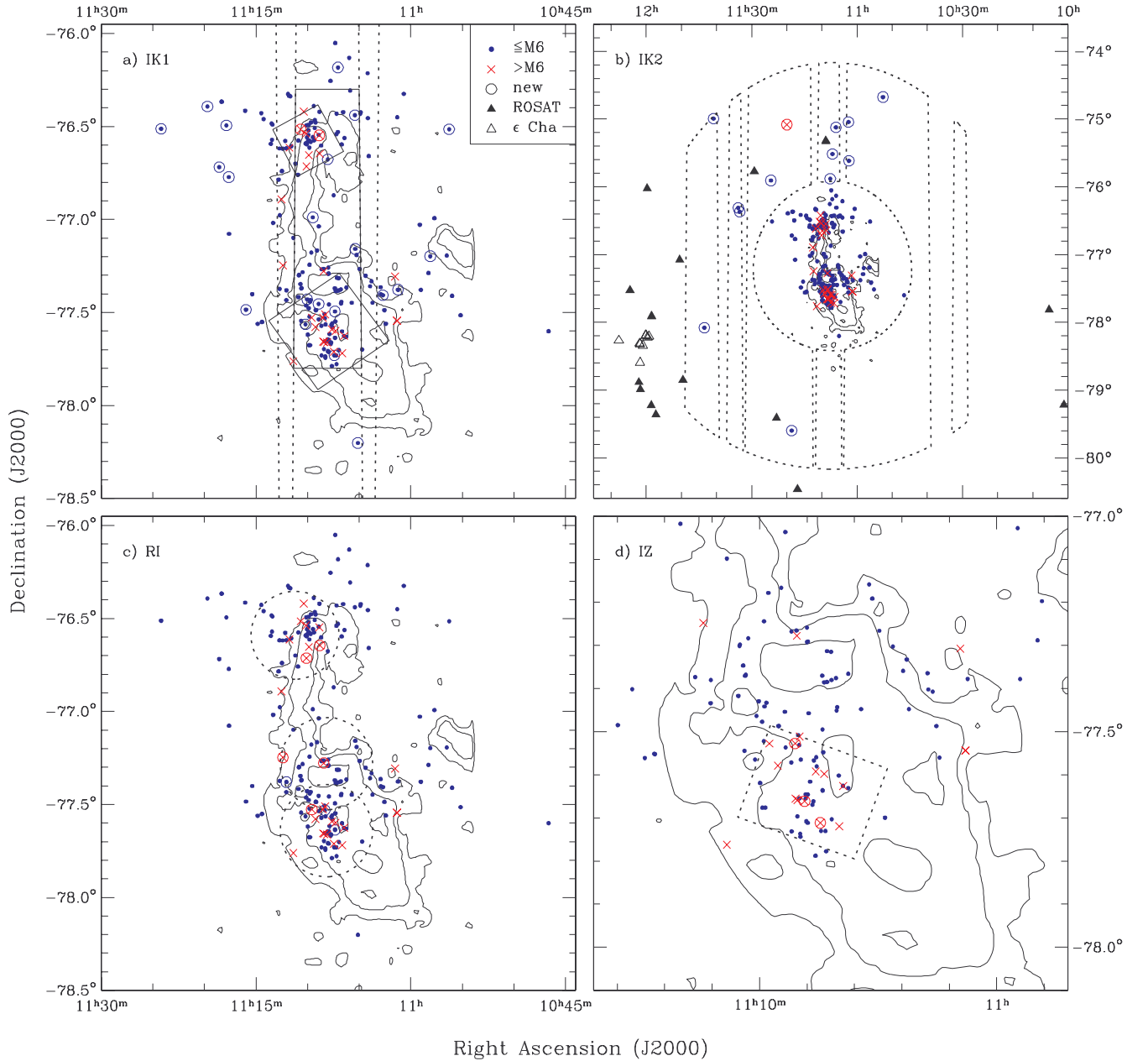}
\caption{
Fields in the Chamaeleon~I star-forming region that have been searched
for new members ({\it dotted lines}). The known members of the cluster 
are indicated ({\it filled circles and crosses}), including the new 
ones found in this work ({\it circles}).
a) In the IK1 survey (see Figure~\ref{fig:ik1}), the central 
$1.5\arcdeg\times0.35\arcdeg$ field was searched most thoroughly for 
new members ({\it large solid rectangle}).
The IMF for this field is shown in Figure~\ref{fig:imf}.
Two narrow strips were not included in this survey because 
the Second DENIS Release contained no data for them ({\it dotted lines}).
For reference, the fields imaged in X-rays by 
{\it Chandra} \cite[][{\it upper solid lines}]{fl04} and {\it XMM} 
\citep[][{\it lower solid lines}]{stel04} are included.
b) The IK2 survey (see Figure~\ref{fig:ik2}) applies to the area between
$r=1.25$-$3\arcdeg$ from the center of Chamaeleon~I for which data are 
available from the Second DENIS Release ({\it dotted lines}).
This map also includes pre-main-sequence stars identified with {\it ROSAT}
\citep[{\it filled triangles},][]{cov97} and probable members of 
the $\epsilon$~Cha young association
\citep[{\it open triangles}, $\tau\sim6$~Myr,][]{fei03,luh04eta}.
c) The RI survey (see Figure~\ref{fig:ri}) applies to three fields imaged 
with IMACS ({\it dotted lines}).
d) The IZ survey (see Figures~\ref{fig:iz} and \ref{fig:ijhk}) applies
to a $0.22\arcdeg\times0.28\arcdeg$ field imaged with ACS ({\it dotted lines}).
The IMF for the ACS field is shown in Figure~\ref{fig:imf}.
The contours represent the extinction map of \citet{cam97} at intervals
of $A_J=0.5$, 1, and 2.
}
\label{fig:map1}
\end{figure}

\begin{figure}
\plotone{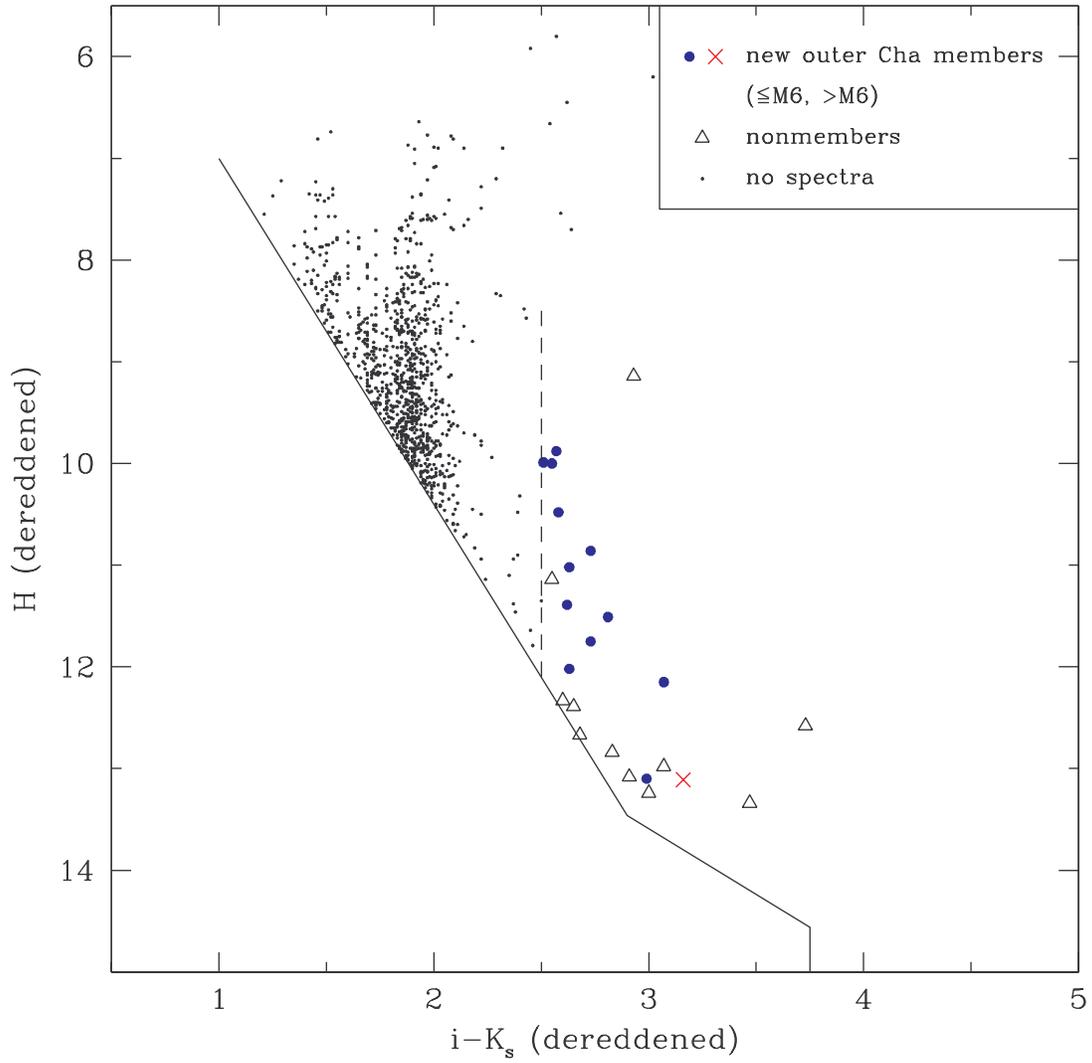}
\caption{
Extinction-corrected color-magnitude diagram for the IK2 survey of
Chamaeleon~I, which applies to the area covered by DENIS in 
Figure~\ref{fig:map1}b. Candidate low-mass members of Chamaeleon~I 
($\gtrsim$M4.5, $M\lesssim0.2$~$M_\odot$) 
were selected based on positions above the solid boundary and right of
the dashed line ($I-K_s>2.5$) and were spectroscopically classified as 
new members ({\it filled circles}) or nonmembers ({\it triangles}).
These measurements are from DENIS ($i$) and 2MASS ($H$, $K_s$).
}
\label{fig:ik2}
\end{figure}

To search for low-mass members of Chamaeleon~I outside of the area of the 
IK1 survey, I again used data from the 2MASS Point Source Catalog and the 
Second DENIS Release. 
For this wide-field survey, named IK2, I considered the area 
between radii of 1.25 and $3\arcdeg$ from 
$\alpha=11^{\rm h}07^{\rm m}00^{\rm s}$, $\delta=-77\arcdeg10\arcmin00\arcsec$
(J2000). The map in Figure~\ref{fig:map1}b indicates the portion of this 
annulus for which data are available from the Second DENIS Release.
As in the IK1 survey, I constructed an extinction-corrected color-magnitude 
diagram from the DENIS and 2MASS data and used it to identify candidate
members of Chamaeleon~I, which are shown on the color-magnitude diagram in 
Figure~\ref{fig:ik2}. 
Unlike IK1, I selected for spectroscopy only the candidates with $I-K_s>2.5$,
which corresponds to spectral types of $\gtrsim$M4.5 and masses of
$M\lesssim0.2$~$M_\odot$. These criteria produced a total of 24 candidate
low-mass stars and brown dwarfs. I have obtained spectra of 23 of these 
candidates (\S~\ref{sec:spec}). The remaining object, 2MASS~J10504888-7829244,
was classified as a galaxy through visual inspection of acquisition images 
obtained during the spectroscopic observations, and thus a spectrum was 
unnecessary. I find that 13 of the targets are pre-main-sequence objects
(\S~\ref{sec:class}).  Kinematic measurements are needed to verify that 
these sources are directly associated with Chamaeleon~I \citep{cov97}, but for 
the purposes of this study, I treat them as members of the star-forming region. 
The positions of these sources are indicated on the map in 
Figure~\ref{fig:map1}b. For comparison, I also show the positions of other 
young stars previously detected within the area encompassed by 
Figure~\ref{fig:map1}b. These additional stars consist of probable members of 
the $\epsilon$~Cha association \citep{fei03,luh04eta} and pre-main-sequence
stars found in the {\it ROSAT} all-sky survey \citep{alc95,alc97,cov97}.

\subsubsection{RI Survey}
\label{sec:ri}

On the night of 2004 January 11, I obtained $R$- and $I$-band images centered at
$\alpha=11^{\rm h}08^{\rm m}10^{\rm s}$ and $\delta=-77\arcdeg38\arcmin00\arcsec$,
$\alpha=11^{\rm h}08^{\rm m}10^{\rm s}$ and $\delta=-77\arcdeg17\arcmin00\arcsec$, and 
$\alpha=11^{\rm h}11^{\rm m}13^{\rm s}$ and $\delta=-76\arcdeg35\arcmin00\arcsec$ (J2000) 
with the Inamori Magellan Areal Camera and Spectrograph
(IMACS) on the Magellan~I telescope at Las Campanas Observatory.
The instrument contained eight $2048\times4096$ CCDs separated by
$\sim10\arcsec$ and arranged in a $4\times2$ mosaic.
The plate scale was $0.202\arcsec$~pixel$^{-1}$ and the unvignetted field of 
view was circular with a diameter of $28\arcmin$. The imaged areas are 
indicated in the map of Chamaeleon~I in Figure~\ref{fig:map1}c.
In each filter, I obtained two dithered images at each exposure time
of 1, 25, and 450~sec.  Photometry and image coordinates
of the sources in these data were measured with DAOFIND and
PHOT under the IRAF package APPHOT.
Aperture photometry was extracted with a radius of six pixels. The background
level was measured in an annulus around each source and subtracted from
the photometry, where the inner radius of the annulus was six pixels and the
width was one pixel.
The photometry was calibrated in the Cousins system by combining
data for standards across a range of colors \citep{lan92} with the
appropriate aperture and airmass corrections. 
Because an atmospheric dispersion corrector was unavailable at the time
of these observations, the image quality varies with position across the 
images. As a result, the photometric uncertainties for these data are
relatively large, with a minimum value of $\sim0.1$~mag. 

As in the IK1 and IK2 color-magnitude diagrams, I defined a boundary
that follows the lower envelope of the sequence of known members in
$I$ versus $R-I$ for separating probable field stars from candidate members
of Chamaeleon~I. I have obtained spectra of 46 of the 116 resulting candidates
(\S~\ref{sec:spec}), which are classified as 38 nonmembers and eight members
(\S~\ref{sec:class}). Figure~\ref{fig:ri} shows the color-magnitude diagram 
of these 46 sources and the previously known members that have 
unsaturated IMACS photometry. The positions of the new members from the RI
survey are indicated on the map in Figure~\ref{fig:map1}c.

\subsubsection{IZ Survey}
\label{sec:iz}

To search for members of Chamaeleon~I at very low masses 
($M\sim0.01$~$M_\odot$), I used deep optical and near-IR images of
the densest portion of the southern subcluster that were obtained by 
\citet{luh05cha}. 
The optical images were collected with the Advanced Camera for Surveys (ACS) 
aboard {\it HST} through the F775W and F850LP filters, which are 
similar to the $i\arcmin$ and $z\arcmin$ filters from the Sloan Digital
Sky Survey \citep{sir05} . These images
covered a $0.22\arcdeg\times0.28\arcdeg$ area, which is indicated in 
Figure~\ref{fig:map1}d. The near-IR imaging was performed with the 
Infrared Side Port Imager (ISPI) at the 4~m Blanco telescope at CTIO through 
$J$, $H$, and $K_s$ filters. These data covered a $20\arcmin\times20\arcmin$ 
area and were centered at $\alpha=11^{\rm h}07^{\rm m}45^{\rm s}$, 
$\delta=-77\arcdeg38\arcmin20\arcsec$ (J2000), completely encompassing the 
ACS field. 

\begin{figure}
\plotone{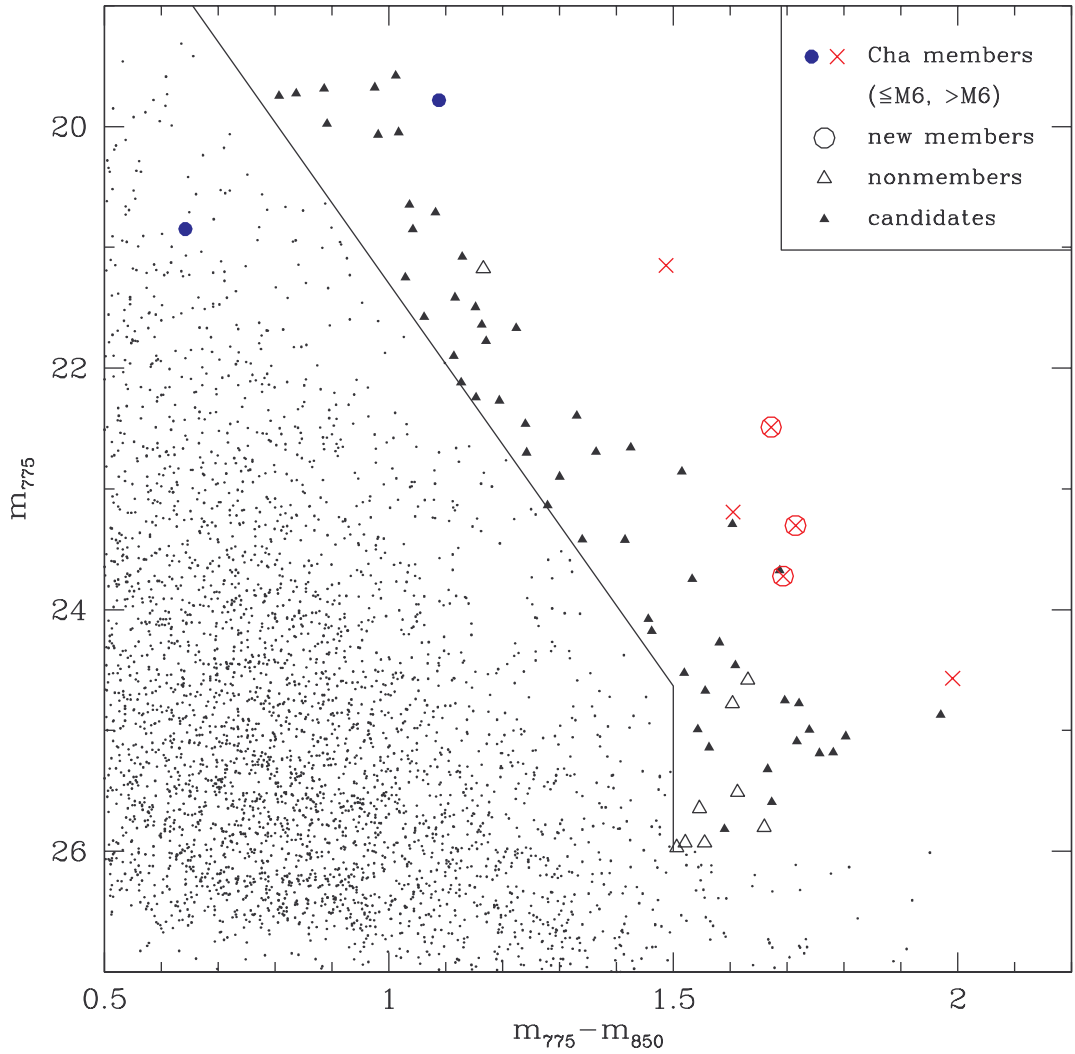}
\caption{
Color-magnitude diagram for the IZ survey of Chamaeleon~I, which applies
to the field imaged with ACS in Figure~\ref{fig:map1}d.
Candidate members of Chamaeleon~I were selected based on positions above 
the solid boundaries in this diagram and in Figure~\ref{fig:ijhk}.
These candidates were spectroscopically classified as 
new members ({\it circles}) or nonmembers ({\it triangles}).
Stars that are candidates members by this diagram but
fall below the solid line in Figure~\ref{fig:ijhk} probably have spectral 
types earlier than M6 and are either field stars or reddened stellar
cluster members ({\it filled triangles}). 
Stars that are too faint to be reliably identified as either field stars or 
candidate members ($m_{775}>26$) have been excluded from this analysis. 
All unsaturated, known members within this field 
are indicated ({\it filled circles and crosses}).
}
\label{fig:iz}
\end{figure}

In Figure~\ref{fig:iz}, I plot a color-magnitude diagram consisting of 
photometry at F751W and F850LP for all unsaturated sources in the ACS images.
The saturation limits of these data correspond to masses that are 
below the hydrogen burning limit for unreddened members of Chamaeleon~I. 
In fact, most of the known substellar members of the cluster ($\gtrsim$M6)
are saturated. Two members with spectral types earlier than M6 are not
saturated because one of them, ISO~79, has significant reddening and the other 
star, Cha J11081938-7731522, is subluminous for its color and spectral type 
and thus may be seen in scattered light (\S~\ref{sec:app1}).
To separate candidate cluster members and probable field stars, 
I defined a boundary below the lower envelope of the sequence of known members.
Because the $I-Z$ colors of late-type field dwarfs do not decrease
significantly from late-M to mid-L \citep{sh00,dah02,dob02},
I have made this boundary vertical at the reddest colors. 
At $m_{775}>26$, I cannot use Figure~\ref{fig:iz} to reliably separate 
field stars and candidate members because the photometric uncertainties 
are too large. Considering only stars at $m_{775}\leq26$, I found 69 
candidates in Figure~\ref{fig:iz}. 

\begin{figure}
\plotone{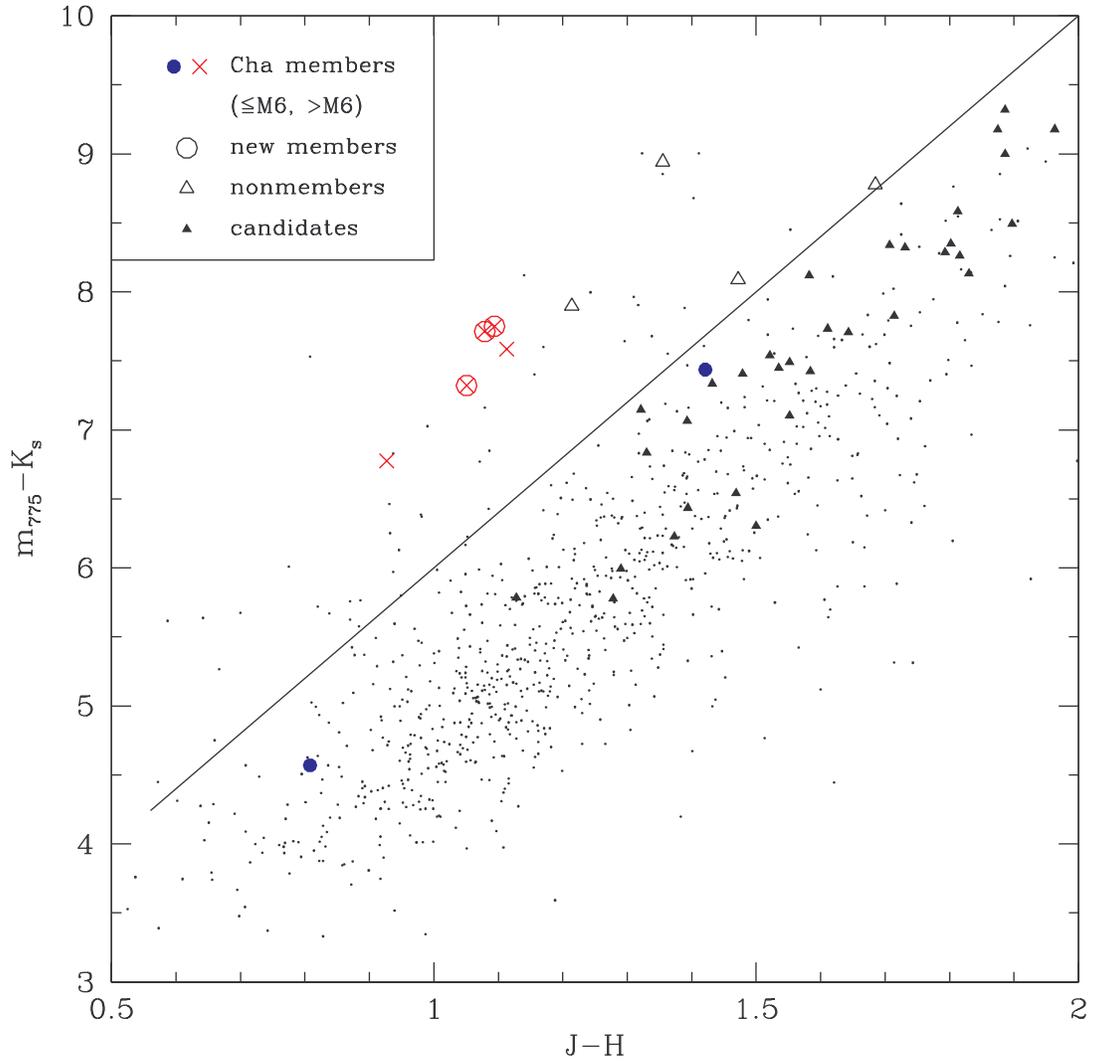}
\caption{
Color-color diagram for the IZ survey of Chamaeleon~I, which applies
to the field imaged with ACS in Figure~\ref{fig:map1}d.
The symbols are the same as in Figure~\ref{fig:iz}.
}
\label{fig:ijhk}
\end{figure}

Color-color diagrams like $I-K$ versus $J-H$ can further refine samples of 
candidate late-type objects found in color-magnitude diagrams \citep{luh00tau}.
Therefore, I constructed an analogous diagram of $m_{775}-K_s$ versus $J-H$ in
Figure~\ref{fig:ijhk} by combining the ACS and ISPI photometry.
To identify objects in this diagram that have colors indicative of late 
spectral types, I have defined a boundary in Figure~\ref{fig:ijhk} 
that approximates the reddening vector for a spectral type of M6, which
corresponds to $M\sim0.1$~$M_\odot$ for members of Chamaeleon~I. 
The slope of this vector was determined from the distribution
of reddened stars in Figure~\ref{fig:ijhk}. The origin of the vector
is placed at $J-H=0.56$ \citep{leg92} and a value of $m_{775}-K_s$ such
that the vector is slightly redder than the color of the known member 
ISO~79 \citep[M5.25,][]{luh04cha}. Objects later than M6 should appear
above this reddening vector. 
Among the 69 candidates from Figure~\ref{fig:iz}, 57 sources have photometry
in $J$, $H$, and $K_s$ and thus are present in Figure~\ref{fig:ijhk}.
Eight of these 57 objects are above the reddening vector, indicating that
they could be later than M6. The 49 candidates below the vector are unlikely
to be substellar members, but some of them could be young stars
with very high reddenings. 
One of the eight late-type candidates, Cha~110913-773444, also exhibits 
mid-IR excess emission in images from the {\it Spitzer Space Telescope}. 
Based on the promising position of this object in Figures~\ref{fig:iz}
and \ref{fig:ijhk} and its mid-IR excess, \citet{luh05cha} obtained a spectrum
of this object and confirmed its youth and late-type nature. 
I have performed spectroscopy on the seven remaining late-type
candidates from Figure~\ref{fig:ijhk} (\S~\ref{sec:spec}), three of which are 
confirmed as low-mass members of Chamaeleon~I (\S~\ref{sec:class}).
I have also obtained spectra of six of the 12 candidates from 
Figure~\ref{fig:iz} that lack $JHK_s$ photometry and thus could not be plotted 
in Figure~\ref{fig:ijhk}. One of these six objects was not detected by ISPI
because it is close to a brighter stellar member, CHXR~73. Spectroscopic
confirmation of this probable companion was presented by \citet{luh06bin}.
The five remaining ACS-only candidates in my spectroscopic sample are 
nonmembers (\S~\ref{sec:class}).

\subsection{Spectroscopy}
\label{sec:spec}

I performed optical and near-IR spectroscopy on 193 candidate members of 
Chamaeleon~I that were identified in \S~\ref{sec:cand}
and 53 miscellaneous targets that are described in this section. 
Some of these objects were observed multiple times. 
Table~\ref{tab:log} summarizes the dates, telescopes, and 
instrument configurations for these observations.
In this section, I provide additional details of the target selection
and observations.
The identity of the sources in each of the categories of targets
described below can be found in Tables~\ref{tab:new} and \ref{tab:non}, 
which list the basis of selection for each target.

I observed one of the IK1 candidates with the Magellan Inamori Kyocera 
Echelle (MIKE). The details of the observations are the same as those 
described by \citet{muz05} for other targets observed on that same night. 
During the IMACS observations in January of 2004, I obtained long-slit and
multi-object spectra of 123 sources, consisting of 85 IK1 candidates,
19 known members of Chamaeleon~I that I have not previously classified, 
four candidate members from \citet{lm04}, six candidate brown dwarfs 
from 2MASS data, two objects with uncertain previous classifications that 
were noted by \citet{luh04cha}, six X-ray sources from \citet{fl04}, and
a candidate companion to the binary ISO~250 \citep{lm04}.
The 2MASS brown dwarf candidates were selected to have 
uncertainties less than 0.1~mag in $J$, $H$, and $K_s$, $0.5\leq J-H\leq1$,
$H-K_s>0.3$, and positions in the central $1.5\arcdeg\times0.35\arcdeg$ area 
indicated in Figure~\ref{fig:map1}a.
The 2MASS candidates, X-ray sources, stars with uncertain 
classifications, and the candidate companion are all classified as nonmembers
in \S~\ref{sec:class}.

During the observations with the Low Dispersion Survey Spectrograph (LDSS-2) 
on the Magellan~II telescope at Las Campanas Observatory on the night of 2004 
March 31, 
I obtained long-slit spectra of 18 and 23 candidates from the IK1 and IK2
surveys, respectively. The one remaining IK2 candidate was not observed
spectroscopically because it was identified as a galaxy on acquisition images
obtained with LDSS-2, as discussed in \S~\ref{sec:ik2}.
On the nights of 2004 April 26 and 27, I performed long-slit spectroscopy
on eight IK1 candidates and multi-slit spectroscopy on 37 RI candidates. 
One of these IK1 candidates was also observed during an earlier run.
During the IMACS observations in January of 2005, I obtained long-slit and
multi-slit spectra of 32 targets, consisting 
of one IK1 candidate, nine RI candidates, a candidate companion to
KG~102 \citep{per05}, one star with an uncertain previous classification, 
eight candidates identified with mid-IR photometry from the {\it Spitzer 
Space Telescope}, five known members that I have not previously classified,
and seven targets observed in earlier runs that needed better signal-to-noise.
Although it lacks previous spectroscopic confirmation of youth and membership,
I am counting the companion T39B as one of the known members.
To provide sufficiently high spectral resolution for the measurement of Li
absorption in this possible companion, it was observed with the 600~l~mm$^{-1}$
grism, which provided coverage of 0.6-0.92~\micron\ and a resolution of 
R=5000. All other targets were observed with the instrument configuration 
that is indicated in Table~\ref{tab:log}.
The {\it Spitzer} candidates were identified with data from \citet{luh05frac}
using the color criteria described in that study. 
Finally, on the nights of 2005 March 23-25, I obtained near-IR spectra of 
12 IZ candidates with the Gemini Near-Infrared Spectrograph (GNIRS) at 
Gemini South Observatory through program GS-2005A-C-13. 
The observing and analysis procedures were 
the same as those described by \citet{luh04ots} for spectroscopy of OTS~44.

\subsection{Classification of Candidate Members}
\label{sec:class}

To measure spectral types and assess membership in Chamaeleon~I for 
the 246 targets in my spectroscopic sample, I applied the classification
methods developed in my previous studies of this kind
\citep{luh99,luh04cha,luh06tau,luh03ic}.
In brief, spectral types were measured from the optical spectra
by comparing them to the averages of dwarfs and giants and to spectra
of previously-classified members of Chamaeleon~I \citep{luh04cha},
while the IR spectra were classified through comparison to IR data
for optically-classified young objects \citep{luh04ots}.
To evaluate the membership of each target, I employed various 
diagnostics based on the spectra, photometry, and other published data,
which included gravity-sensitive lines, emission lines, IR excess emission,
Li absorption, reddening, and radial velocities.

I have classified 50 targets as new members of Chamaeleon~I and 163 targets
as nonmembers. The remaining 33 objects in my spectroscopic sample are 
previously known members. 
The new members, nonmembers, and previously known members are listed
in Tables~\ref{tab:new}, \ref{tab:non}, and \ref{tab:old}, respectively.
2M~J1101-7732 is counted as a previously known member and is omitted from
Table~\ref{tab:old} because it was already presented by \citet{luh04bin},
but the photometry used for its selection is included in Table~\ref{tab:phot}.
Tables~\ref{tab:new}-\ref{tab:old} provide the source identifications, 
the available spectral types, the basis for each object's selection for
spectroscopy, the night on which it was observed, and its evidence of 
membership or nonmembership.
The references for the membership data consist of \citet{luh04cha}, 
\citet{com04}, \citet{luh05frac}, this work, and unpublished {\it Spitzer} data.
For sources that are not in the 2MASS Point Source Catalog, I have assigned new 
coordinate-based names. Additional names for the previously known members 
have been compiled by \citet{car02} and \citet{luh04cha}.
All spectral types in Table~\ref{tab:new} are from this work. 
The spectra of the X-ray-selected targets were described by \citet{fl04}. 
Comments on the classifications of other individual sources are provided 
in \S~\ref{sec:app1}.

\begin{figure}
\epsscale{1.2}
\plotone{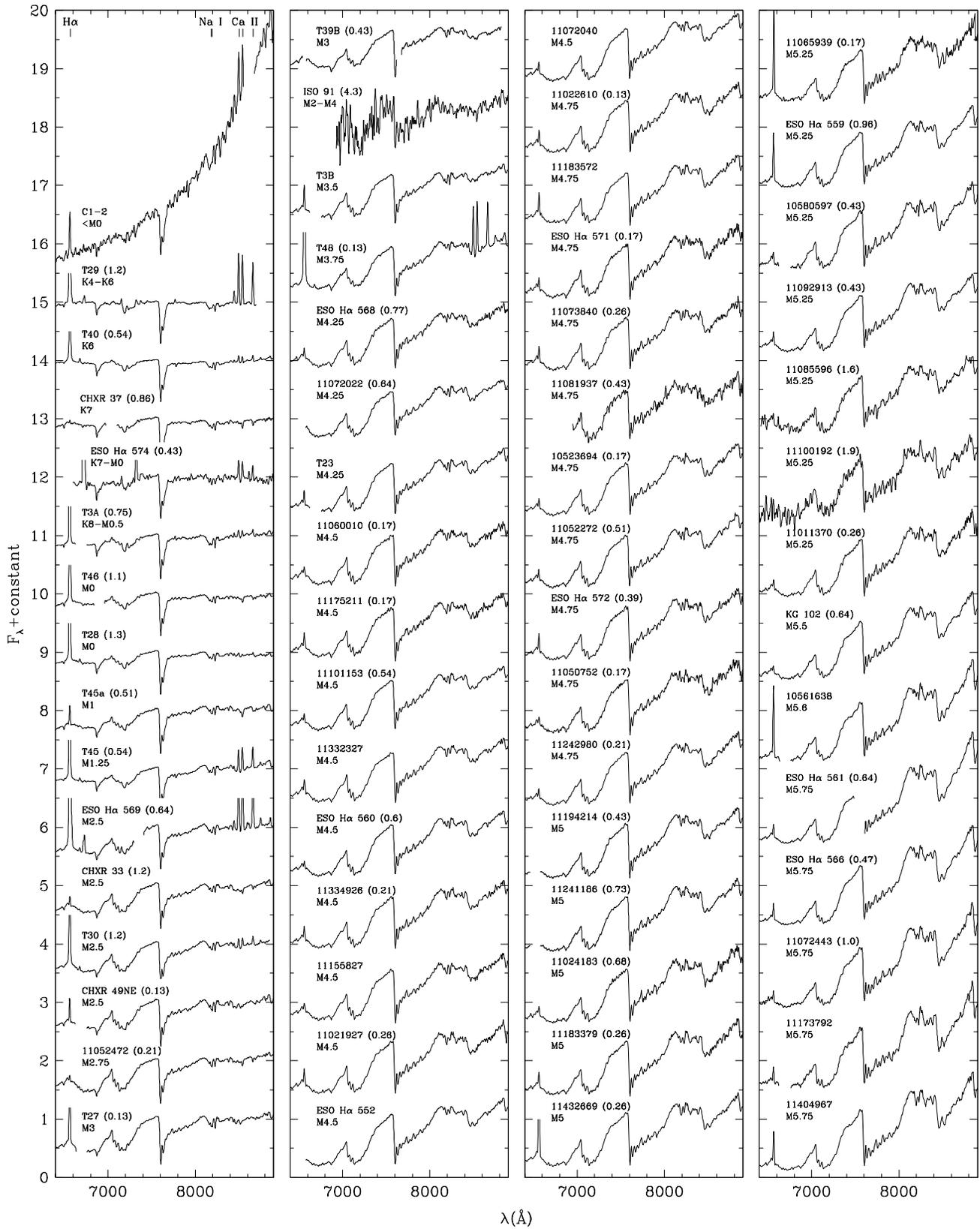}
\caption{
Optical spectra of previously known members of Chamaeleon~I 
and new members identified in this work.
With the exception of C1-2, the spectra have been corrected for extinction, 
which is quantified in parentheses by the magnitude difference of the reddening
between 0.6 and 0.9~\micron\ ($E(0.6-0.9)$).
The data are displayed at a resolution of 13~\AA\ and are normalized at
7500~\AA.
}
\label{fig:op1}
\end{figure}

\begin{figure}
\epsscale{1.2}
\plotone{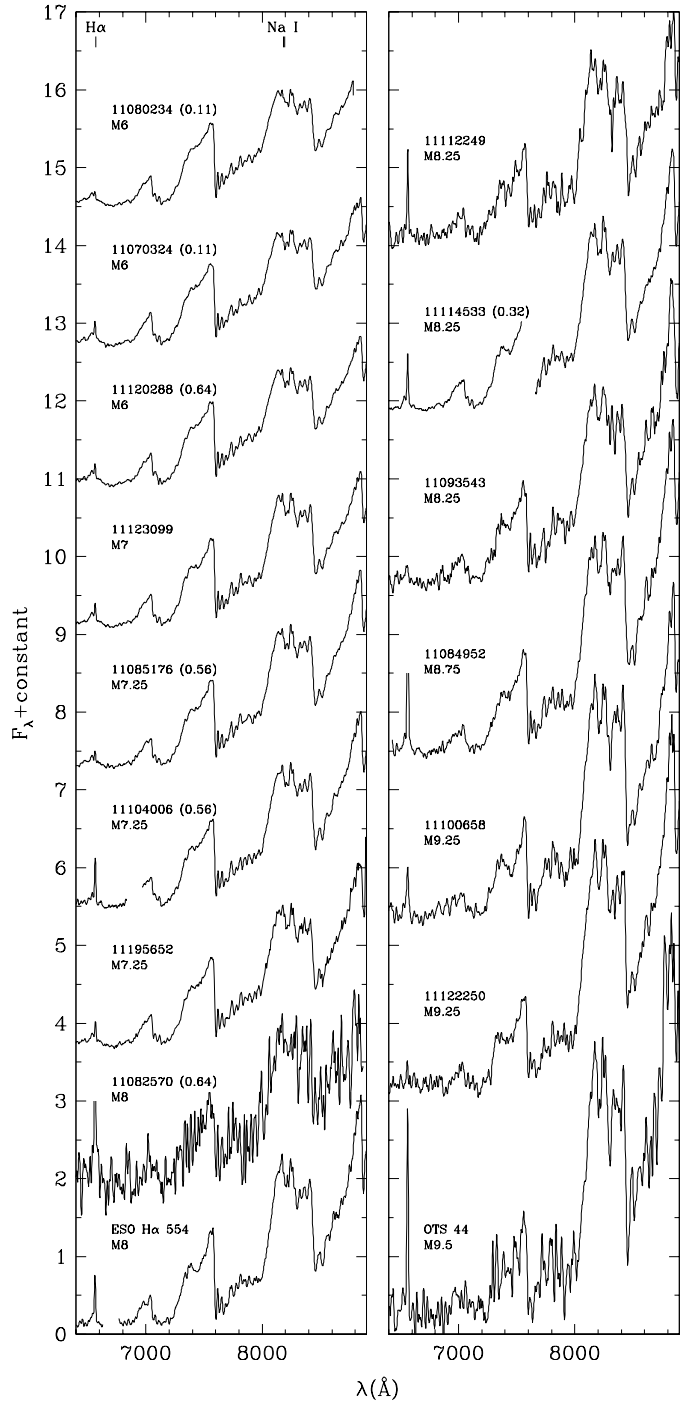}
\caption{
More optical spectra of previously known and new members of Chamaeleon~I
(see Fig.~\ref{fig:op1}).
}
\label{fig:op2}
\end{figure}

\begin{figure}
\epsscale{0.7}
\plotone{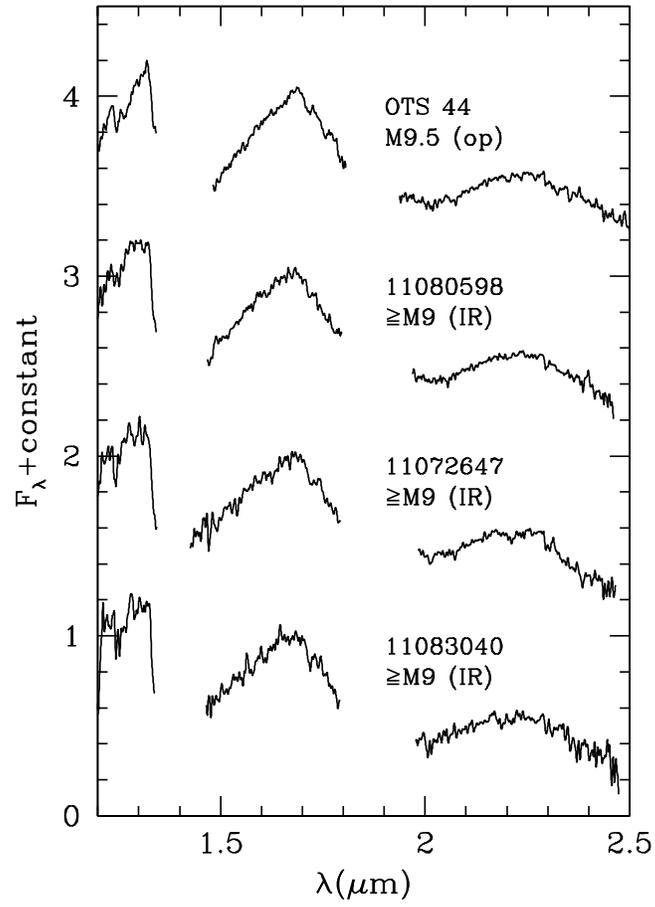}
\caption{
Near-IR spectra of the three new members of Chamaeleon~I from
Figures~\ref{fig:iz} and \ref{fig:ijhk} ({\it eight digit identifications}) 
and a previously known member, OTS~44.  The former are 
labeled with the spectra types derived from this comparison to 
OTS~44, which has an optical type of M9.5 (Figure~\ref{fig:op2}), and
other optically-classified late-type members.
The spectra are displayed at a resolution of $R=500$, are normalized at 
1.68~\micron, and are dereddened (\S~\ref{sec:class}).
}
\label{fig:ir}
\end{figure}

The optical and IR spectra of the new and previously-known members of 
Chamaeleon~I in my spectroscopic sample are shown in
Figures~\ref{fig:op1}-\ref{fig:ir}. 2M~J1106-7732 is excluded because 
\citet{luh04bin} already presented its spectrum and C1-25 is not shown 
because its spectrum has very low signal-to-noise. Some of the spectra obtained
with IMACS exhibit gaps in wavelength coverage that are produced by the 
boundaries between adjacent CCDs.

\subsection{Completeness of Census}
\label{sec:complete}

In \S~\ref{sec:imf}, I will construct IMFs for a
$1.5\arcdeg\times0.35\arcdeg$ field encompassing most of the Chamaeleon~I 
cloud (Figure~\ref{fig:map1}a) and the $0.22\arcdeg\times0.28\arcdeg$ field 
in the southern subcluster that was imaged with ACS (Figure~\ref{fig:map1}d).
To ensure that these IMFs are accurate representations of the stellar 
population in Chamaeleon~I, in this section I evaluate the completeness of 
the current census of known members within these two fields as a function 
of mass and extinction. I also perform a completeness analysis for the 
substellar members within $3\arcdeg$ of the center of Chamaeleon~I, the
results of which will be used in \S~\ref{sec:spatial} to measure
the large-scale spatial distribution of brown dwarfs. 

\begin{figure}
\plotone{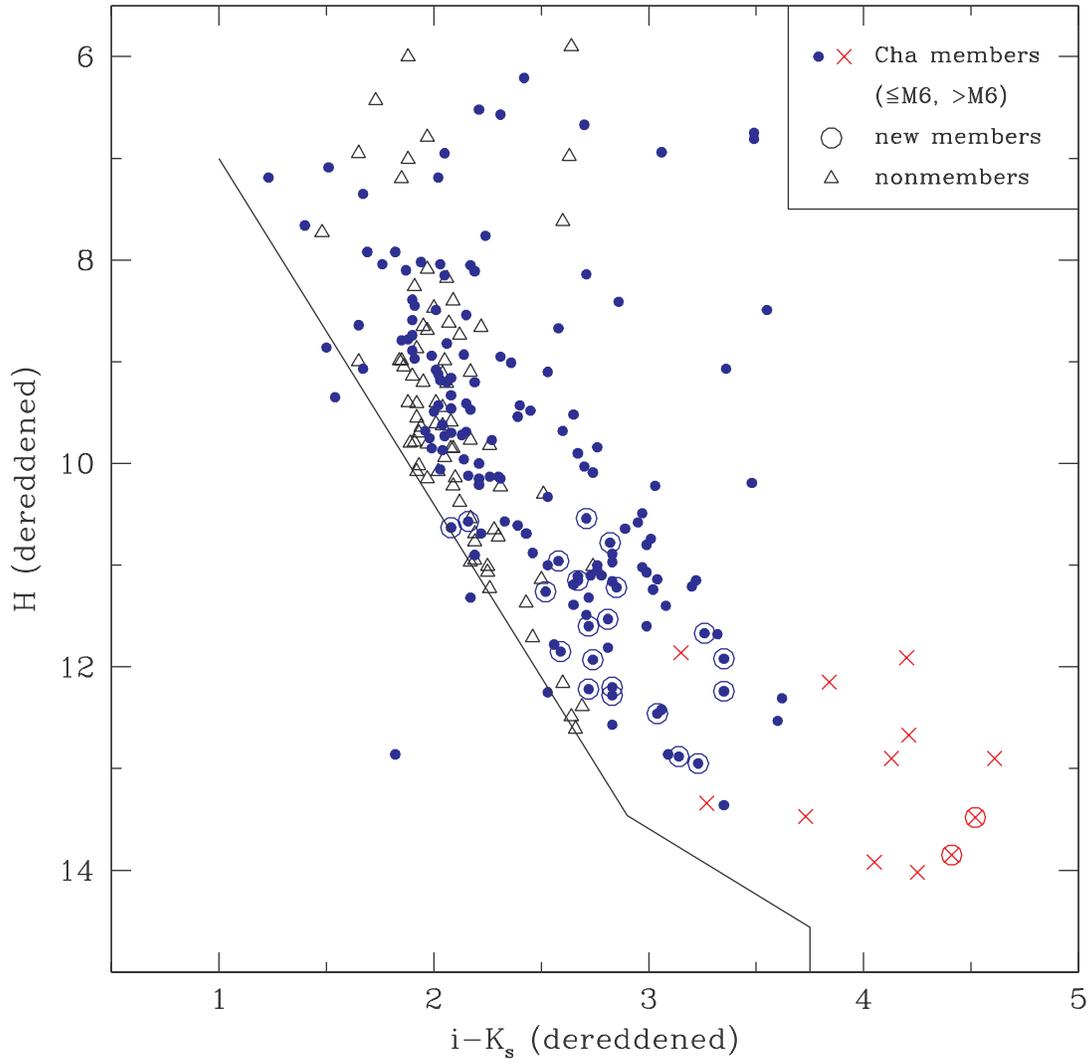}
\caption{
Extinction-corrected color-magnitude diagram for the IK1 survey of
Chamaeleon~I, which applies to the area covered by DENIS in
Figure~\ref{fig:map1}a. Candidate members of Chamaeleon~I were selected 
based on positions above the solid boundary and were spectroscopically 
classified as new members ({\it circles}) or nonmembers ({\it triangles}).
All known members within this field are indicated 
({\it filled circles and crosses}).
These measurements are from DENIS ($i$) and 2MASS ($H$, $K_s$).
}
\label{fig:ik1}
\end{figure}

I first consider the $1.5\arcdeg\times0.35\arcdeg$ field. 
In Figure~\ref{fig:jh1}, I show the diagram of $J-H$ versus $H$ from 2MASS
for this area.
The completeness limits of the 2MASS photometry are taken to be the magnitudes
at which the logarithm of the number of sources as a function of magnitude
departs from a linear slope and begins to turn over.
In Figure~\ref{fig:jh1}, I have plotted the 10~Myr isochrone from \citet{bar98}
for $A_J=0$ and 1.2. I selected this isochrone because most members of 
Chamaeleon~I fall above it on the Hertzsprung-Russell (H-R) diagram
(\S~\ref{sec:hr}). A comparison of the 2MASS completeness limits to the 
isochrones indicates that the photometric data should be 
complete for members with $M/M_{\odot}\geq0.03$ ($\lesssim$M8) and $A_J\leq1.2$,
with the exception of close companions and objects that are simultaneously
near $M/M_{\odot}\sim0.03$, $A_J\sim1.2$, and $\tau\sim10$~Myr. 
I have omitted in Figure~\ref{fig:jh1} the 2MASS sources that are nonmembers
according to spectroscopy or that are likely to be field stars 
because they are below the sequences of known members 
in Figures~\ref{fig:ik1} and \ref{fig:ri}.
The remaining sources either have been confirmed as members or lack sufficient
data for the assessment of their membership.
Based on these data, the current census in the
$1.5\arcdeg\times0.35\arcdeg$ field should be nearly 100\% complete 
for $M/M_{\odot}\geq0.03$ and $A_J\leq1.2$.
These completeness limits do not apply to objects that are seen in scattered 
light (e.g., edge-on disks) because they often appear below cluster
sequences on color-magnitude diagrams and thus can be mistaken for field stars. 

My ISPI near-IR images of Chamaeleon~I encompassed the entire 
$0.22\arcdeg\times0.28\arcdeg$ ACS field and are deeper than 2MASS.
Therefore, for this field I have constructed the diagram of $J-H$ versus $H$ in
Figure~\ref{fig:jh2} from a combination of 2MASS and ISPI photometry. 
For sources that are present in both sets of data, I adopt the measurement from
ISPI if the photometric uncertainty from 2MASS is larger than 0.05~mag.
The completeness limit of the ISPI data is estimated in the same manner as
for 2MASS earlier in this section. 
As with Figure~\ref{fig:jh1}, I show the 10~Myr isochrone from
\citet{bar98}, except for a higher extinction of $A_J=1.4$
and a lower minimum mass of 0.015~$M_\odot$. 
Sources that are likely to be nonmembers based on spectroscopy or their
positions in Figures~\ref{fig:ik1}, \ref{fig:ri}, and \ref{fig:iz} are 
omitted from Figure~\ref{fig:jh2}. I also reject sources that are 
resolved as galaxies in the ACS images. 
The remaining IR sources consist of confirmed members and sources with 
undetermined membership. A comparison of these two populations to the
reddened isochrone and the ISPI completeness limit demonstrates that
the current census of members within the ACS field is complete for
$M/M_{\odot}\geq0.01$ and $A_J\leq1.4$.

\begin{figure}
\plotone{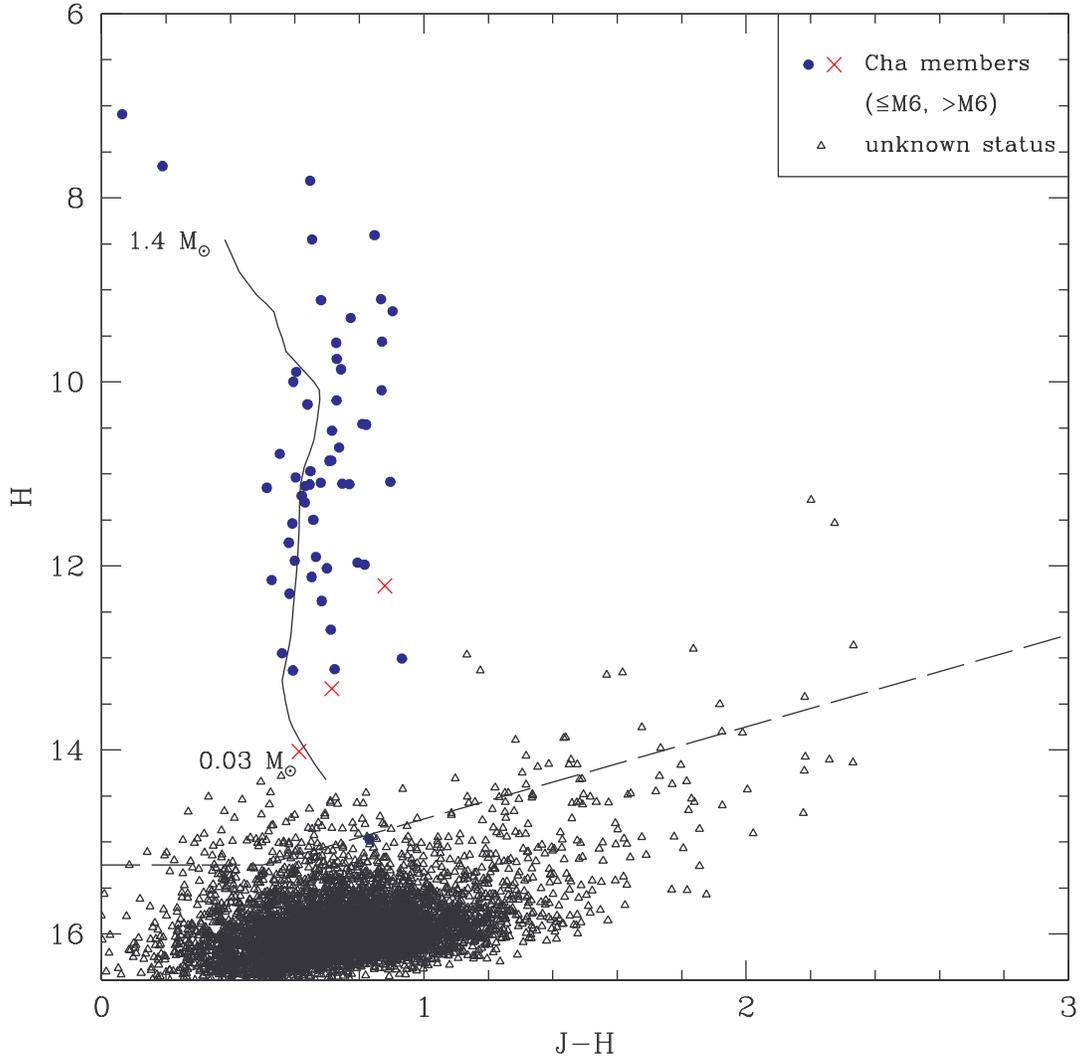}
\caption{
$J-H$ versus $H$ from 2MASS for the area outside of the 
$1.5\arcdeg\times0.35\arcdeg$ field in Figure~\ref{fig:map1}a and within
a radius of $3\arcdeg$ from the center of Chamaeleon~I for which data are
available from the Second DENIS Release 
(see Figure~\ref{fig:map1}b).
I have omitted the stars classified as nonmembers through spectroscopy
(Tables~\ref{tab:non}) as well as objects that are probable field stars
because they are below the sequences of known members
in Figures~\ref{fig:ik1} and \ref{fig:ik2}. 
I have also excluded ones with colors too blue to be substellar members
of the cluster ($i-K_s<3$, $H-K_s<0.35$), with the exception of known members.
The remaining sources shown in this diagram consist of known members 
of Chamaeleon~I ({\it filled circles and crosses}) and stars whose membership 
is undetermined ({\it triangles}). 
The 10~Myr isochrone (1.4-0.03~$M_{\odot}$) from the evolutionary models of
\citet{bar98} is shown with no extinction ({\it solid line}).
These measurements have completeness limits of $J=15.75$ and $H=15.25$
({\it long dashed line}).
}
\label{fig:jh3}
\end{figure}

Finally, I evaluate the completeness of the current census of brown dwarfs 
for the area outside of the $1.5\arcdeg\times0.35\arcdeg$ field 
and within a radius of $3\arcdeg$ from the center of Chamaeleon~I. 
I consider only the regions that are covered by the Second DENIS Release,
which are indicated in Figure~\ref{fig:map1}.
The diagram of $J-H$ versus $H$ from 2MASS for this area is shown in 
Figure~\ref{fig:jh3}.
I have excluded 2MASS sources that have been spectroscopically classified
as nonmembers or that are likely to be field stars because they are below 
the sequences of known members in Figures~\ref{fig:ik1} and \ref{fig:ik2}.
Because I am examining the completeness of only brown dwarfs in Chamaeleon~I
and not stellar members for the area in question, I have also omitted
2MASS sources whose colors are too blue for brown dwarfs; nearly all 
known members of Chamaeleon~I with spectral types later than M6 have
$i-K_s>3$ and $H-K_s>0.35$, so I have applied these thresholds. 
The resulting color-magnitude diagram in Figure~\ref{fig:jh3} contains 
known members and sources whose membership is undetermined. 
Because most of the area considered here is outside of the Chamaeleon~I
dark clouds, the known members within it exhibit little extinction ($A_J<0.6$)
and the same is expected for any undiscovered members. For these low levels
of extinction, Figure~\ref{fig:jh3} indicates that the current census of 
Chamaeleon~I is complete for $M/M_{\odot}\geq0.03$ ($\lesssim$M8) out to a 
radius of $3\arcdeg$ for the area covered by DENIS and outside of the central
$1.5\arcdeg\times0.35\arcdeg$ field. 

\subsection{Summary of Updated Census}

It is useful to describe the statistics of the updated census of Chamaeleon~I 
that includes the new members found in this work. 
The census presented by \citet{luh04cha} consisted of 158 members, eight 
of which were later than M6. SGR1 was included in that list of members, but I
now classify it as a field star (\S~\ref{sec:app1}) and remove it from the
census. During the time since \citet{luh04cha}, 16 members have been identified
by \citet{luh04bin}, \citet{luh04ots,luh05cha,luh06bin}, and \citet{com04}.
I also add ESO~H$\alpha$~281 from \citet{bz97}, which was
overlooked by \citet{luh04cha}.
T3B and T39B were previously believed to be companions but were not
counted as members by \citet{luh04cha} because they lacked spectroscopic
confirmation. Based on my spectral classifications of T3B and T39B, 
I include them in the census of members. 
When the 50 new members from this work are included, the 
latest census of Chamaeleon~I contains 226 known members, 28 of which are
later than M6\footnote{The number of objects in a given range of spectral
types depends on the measurements that are adopted. I am adopting
spectral types from this work, \citet{luh04cha,luh04bin}, and
\citet{luh04ots,luh05cha,luh06bin}.}.

\section{Star Formation History}
\label{sec:hr}

\subsection{H-R Diagram}

To place the members of Chamaeleon~I on the H-R diagram, I first estimate
their extinctions, effective temperatures, and bolometric luminosities.
I exclude from this analysis 11 of the 226 known members whose 
spectral types are too uncertain for such estimates. 
When constructing an H-R diagram from an earlier census of Chamaeleon~I,
\citet{luh04cha} adopted the average of the extinctions estimated from
optical spectra and near-IR colors when both were available. 
I use a different approach in this work, adopting only the 
spectroscopically-measured extinctions when possible. 
Following \citet{luh04cha}, the reddenings of the optical spectra are 
quantified by the color excess between 0.6 and 0.9~\micron, which is denoted as 
$E(0.6-0.9)$ in Figures~\ref{fig:op1} and \ref{fig:op2}. These excesses are
converted to $A_J$ with the relation $A_J=E(0.6-0.9)/0.95$ \citep{luh04cha}.
Extinction estimates from spectra are available for 198 members
\citep[\S~\ref{sec:new},][]{luh04cha,luh04bin,luh04ots,luh06bin}.
For the remaining 17 stars, I estimate extinctions from $J-H$ and $H-K_s$
in the manner described by \citet{luh04cha}.

The uncertainties of the extinctions estimated from spectroscopy 
are determined by the errors in relative flux calibration of the spectra. 
All of the targets observed
with long-slit spectroscopy were observed with the slit aligned at the
parallactic angle, which prevents errors in relative flux calibration
due to atmospheric differential refraction \citep{fil82}.
On the other hand, the multi-slit observations generally were not performed
at the parallactic angle because a given slit mask required a specific
position angle. To ensure accurate relative flux calibration, I included
slits in each mask for known members that have been previously observed with 
long-slit spectroscopy. I then compared the spectral slopes of the 
multi-slit and long-slit data to check the accuracy of the former. 
This analysis indicates errors in the relative flux calibration of the
multi-slit data that correspond to $A_J\sim0.13$. 

Spectral types are converted to effective temperatures and bolometric
luminosities are derived from $J$-band magnitudes with the same
temperature scale and bolometric corrections used by \citet{luh04cha}, 
except that I adopt the $J$-band bolometric corrections from \citet{dah02}
for spectral types of M6 and later. 
For computing the luminosities, I adopt a distance modulus of 6.05
\citep{luh07cha}, which differs slightly from the value of 6.13 used 
in \citet{luh04cha}.
For most members of Chamaeleon~I, I used $J$-band measurements from 
2MASS or my ISPI images. However, T39B, 2MASS~11072022-7738111, 
ESO~H$\alpha$~281, T3B, T33B, and CHXR~73B are blended with brighter
stars in those data. 
I estimate near-IR magnitudes for the first four sources by 
applying the differential photometry from \citet{che95} and \citet{cor06}
to the 2MASS data. For T33A/B and CHXR~73B, I adopt the resolved photometry
from \citet{hai04} and \citet{luh06bin}, respectively.
The typical uncertainties in $A_J$, $J$, and BC$_J$ ($\sigma\sim0.13$, 0.03, 
0.1) correspond to errors of $\pm0.07$ in the relative values of 
log~$L_{\rm bol}$. When an uncertainty in the distance modulus is included 
($\sigma\sim0.13$), the total uncertainties are $\pm0.08$.

The extinctions, effective temperatures, bolometric luminosities, and adopted
spectral types for the 215 known members of Chamaeleon~I with spectral 
classifications are listed in Table~\ref{tab:hr}.
These temperatures and luminosities are plotted in Figure~\ref{fig:hr}
for members later than G5. Separate H-R diagrams are shown for the members
north and south of $\delta=-77\arcdeg$, which is the approximate midpoint
between the northern and southern subclusters.
The H-R diagram for types earlier than G5 is the same as the one in 
\citet{luh04cha}, except for a small offset in luminosities because 
of the different adopted distances.

\subsection{Distribution of Ages}

\begin{figure}
\plotone{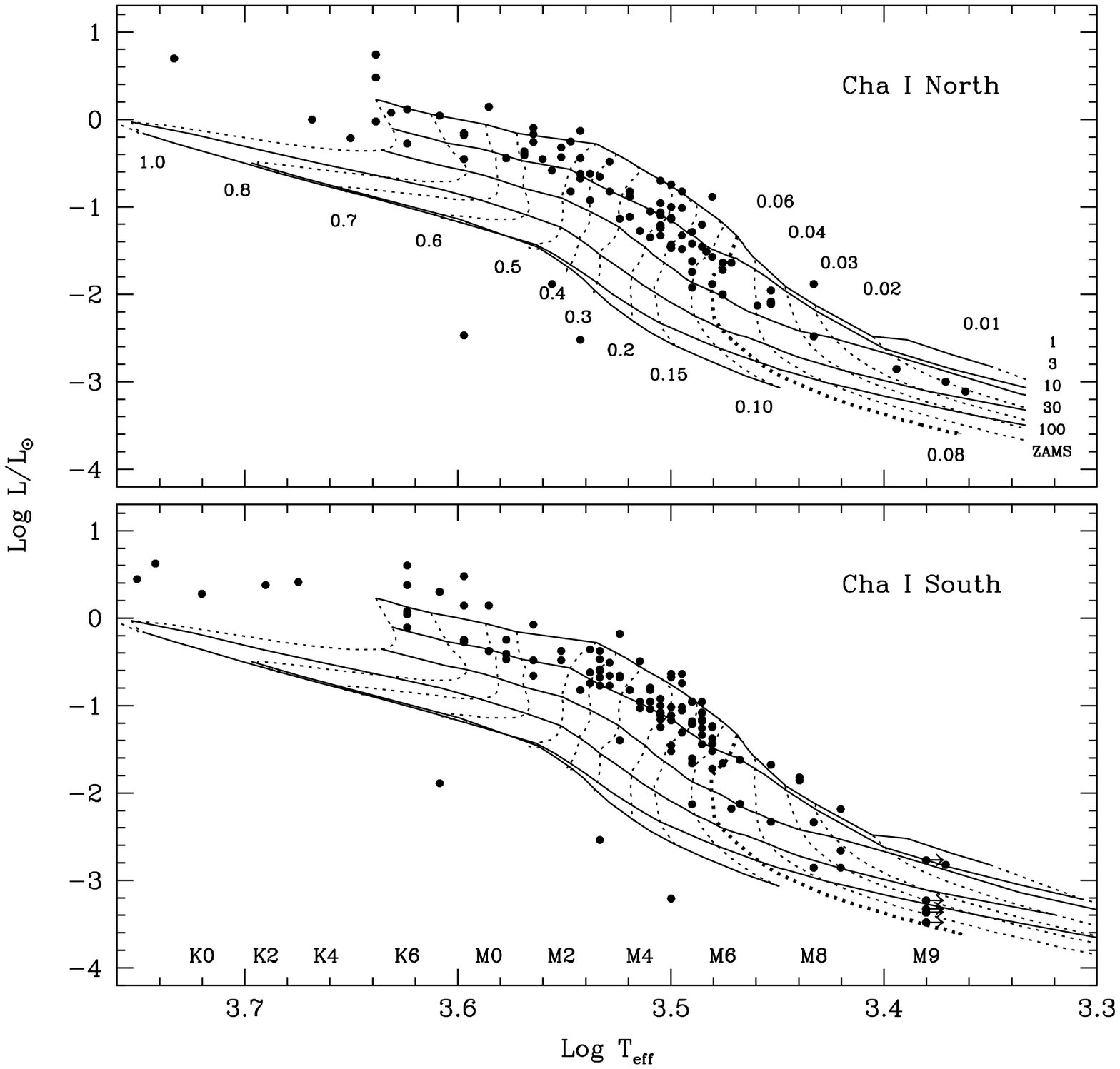}
\caption{
H-R diagrams for the northern ({\it top}, $\delta>-77\arcdeg$) and
southern ({\it bottom}, $\delta<-77\arcdeg$) subclusters in Chamaeleon~I.
These data are shown with the theoretical evolutionary models of
\citet{bar98} ($0.1<M/M_\odot\leq1$) and \citet{cha00} ($M/M_\odot\leq0.1$),
where the mass tracks ({\it dotted lines}) and isochrones ({\it solid lines})
are labeled in units of $M_\odot$ and Myr, respectively.
The typical uncertainties are $\pm0.25$~subclass and $\pm0.08$ for 
spectral types and log~$L_{\rm bol}$, respectively.
}
\label{fig:hr}
\end{figure}

\begin{figure}
\plotone{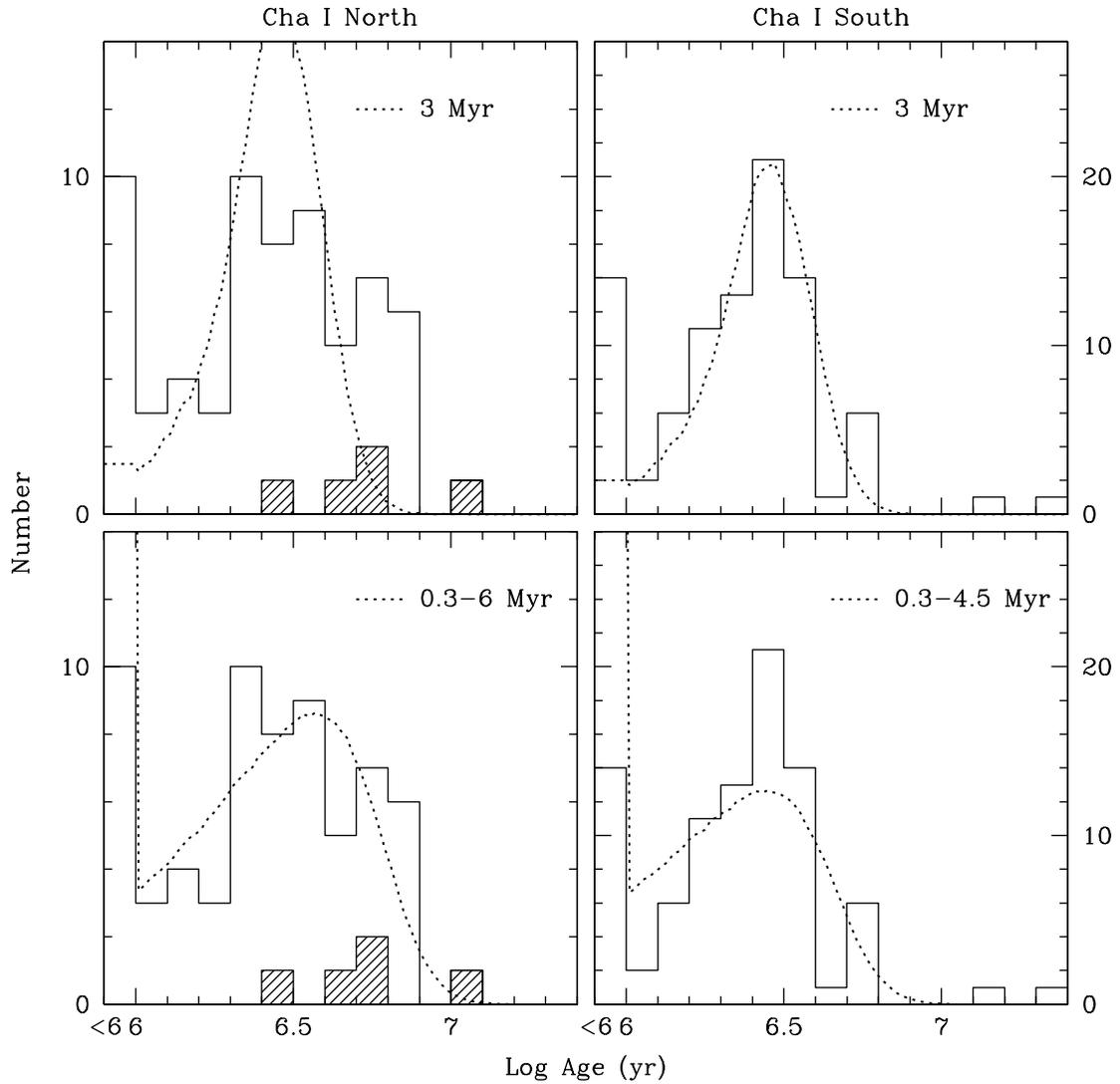}
\caption{
Distributions of isochronal ages from Figure~\ref{fig:hr} for stars between 
0.1 and 1~$M_\odot$ in the northern ({\it left}, $\delta>-77\arcdeg$) and 
southern ({\it right}, $\delta<-77\arcdeg$) subclusters in Chamaeleon~I.
For comparison, I show the distributions from a burst of star formation 
({\it upper dotted lines}) and an extended period of constant star formation 
({\it lower dotted lines}) convolved with the expected error distributions
and normalized to the data at 1-10~Myr.
The five young stars directly north of the Chamaeleon~I clouds 
in Figure~\ref{fig:map1}b ({\it shaded histogram}, $\delta>-76\arcdeg$) 
appear to have older ages, on average, than the two subclusters. 
}
\label{fig:ages}
\end{figure}

In the H-R diagram for Chamaeleon~I from \citet{luh04cha}, the sequence for
Chamaeleon~I closely resembled that of IC~348 in terms of age, with both
clusters exhibiting median ages of $\sim2$~Myr with the models of 
\citet{bar98} and \citet{cha00}.
The same result is produced by the updated data shown in Figure~\ref{fig:hr}.
Because the census of Chamaeleon~I is now larger and more complete, 
I can compare the ages of the northern and southern subclusters. 
In Figure~\ref{fig:hr}, the northern sequence appears to be slightly 
fainter, and hence older, than the southern one. To better illustrate
this age difference, I compare the distributions of model ages for the 
two subclusters in Figure~\ref{fig:ages}. For masses of 0.1 to 1~$M_\odot$, 
the northern and southern subclusters have median ages of 2.8 and 2.5~Myr,
respectively.
For masses below 0.1~$M_\odot$, each sequence in Chamaeleon~I becomes
somewhat older on the model isochrones. Rather than a true variation of 
age with mass, this feature may be caused by an error in the adopted 
temperature scale or evolutionary models.

To interpret the distributions of isochronal ages in Figure~\ref{fig:ages}
in terms of star formation histories, I must account for the observational
errors in the age estimates.
To do this, I assume that the errors follow Gaussian distributions.
I adopt $\sigma=0.13$, 0.1, and 0.1~mag to represent the
errors from extinction ($A_J$), variability, and bolometric corrections.
The sum of these errors in quadrature corresponds to $\sigma$(log~$L)=0.08$,
which is approximately equivalent to $\sigma$(log~$t)=0.12$ \citep{har01}.
Because the depth of Chamaeleon~I should be only a few pc, the uncertainties 
in relative distances should have negligible
effect on the spread in luminosities. I also account for binaries in the 
manner described by \citet{har01}, except that I assume an unresolved binary 
fraction of 30\%, which should be a reasonable value for a sample like the one
in Chamaeleon~I that is dominated by low-mass stars \citep{rg97}.

I now convolve the error distributions with model star formation histories
and compare the results to the distributions of isochronal ages in 
Figure~\ref{fig:ages}.
I consider two types of star formation histories, an instantaneous burst
and an extended period of constant star formation.
For the burst, I select an age that reproduces the median of the 
isochronal ages in Chamaeleon~I. 
As shown in Figure~\ref{fig:ages}, a burst at 3~Myr closely resembles the age 
distribution of the southern sample at ages beyond 1~Myr. The agreement
in the widths of the observed and model distributions suggests that the 
combined observational errors are not significantly larger than the adopted 
value of $\sigma$(log~$t)=0.12$. 
On the other hand, the northern sample exhibits a distribution 
of ages that is broader than the model distribution for a single burst.
If the observational errors are similar between the two halves of Chamaeleon~I,
then the broader age distribution in the north indicates the presence of
a larger age spread, and that star formation began 1-2~Myr earlier 
in the northern subcluster than in the southern one. 

For both the northern and southern subclusters, the burst model produces too 
few stars with isochronal ages less than 1~Myr. Indeed, the presence of 
protostars in this region demonstrates that star formation is ongoing 
\citep{rei96}.
Therefore, I also consider a constant star formation rate that extends to the
present time. For comparison to the data in Figure~\ref{fig:ages},
I select a lower limit of 0.3~Myr for the ages of stars in this model 
because most of the protostars lack spectral classifications and thus 
are absent from the H-R diagram in Figure~\ref{fig:hr}. 
The same approach was used by \citet{har01} for Taurus. 
After varying the upper age limit for this model, I find that values of 
6 and 4.5~Myr provide the best match to the data in the northern and 
southern samples, respectively, although they produce too many stars younger 
than 1 Myr. The models for a burst and periods of constant star formation 
bracket the data for the two subclusters, which suggests that 
star formation has occurred for the past 3-4~Myr in the south and 5-6~Myr in
the north at rates that have declined with time.
Thus, I do not find evidence in these data for Chamaeleon~I to support the
idea that star formation begins slowly and accelerates during    
the lifetime of a molecular cloud \citep{pal97,pal99,pal00,pal02}.

In the preceding analysis, I have inferred ages for stars in Chamaeleon~I
with the model isochrones of \citet{bar98} and \citet{cha00}.
Use of other models would produce different age estimates. 
However, because the evolution of low-mass stars along
Hayashi tracks approximately follows log~$t\propto$~log~$L$ \citep{har98},
the distribution of log~$t$ is a reflection of the distribution of
observed log~$L$ for any set of models. Thus, the shape of the distribution
of isochronal ages shown in Figure~\ref{fig:ages} is insensitive to the 
choice of models, while the average age can vary by 0.1-0.2 in log~$t$
from one set of models to another because of differences in birthlines.

\subsection{Cloud Morphology}
\label{sec:morph}

The difference in isochronal ages between the northern and southern subclusters 
is consistent with the relative levels of obscuration toward their members. 
For instance, the fraction
of stars residing outside of the extinction contours in Figure~\ref{fig:map1}
is higher in the northern subcluster than in the southern one. 
The presence of a north-south age gradient is further supported by
an examination of the string of five young stars directly north of 
the cloud at $\delta>-76\arcdeg$ that have been found in this work 
(see Figure~\ref{fig:map1}). The distribution of the isochronal ages 
of these stars is shown in Figure~\ref{fig:ages}, 
which indicates that these stars are older, on average, than the northern 
and southern samples. Based on these age data, I speculate that star formation 
in this region began in a small cloudlet north of the currently existing 
dark clouds, producing the B star HD~96675 and $\sim5$-10 other stars, and 
propagated south to form the two larger subclusters. 
The apparent gradient in median age and cloud morphology between the northern 
and southern subclusters implies that the Chamaeleon~I clouds are
dissipating on a time scale of $\sim6$~Myr after the initiation of star 
formation, which is consistent with previous estimates of molecular cloud 
lifetimes \citep[][references therein]{elm00,har01b}.

According to theoretical evolutionary models (\S~\ref{sec:masses}), 
the masses of the 215 members with spectral classifications 
range from $\sim0.01$ to 3.5~$M_\odot$ and have a median value of 
0.21~$M_\odot$. The combined mass of these objects is 86~$M_\odot$. 
After including the known members that lack measured spectral types,
the total mass of the known stars and brown dwarfs in Chamaeleon~I is 
$\sim100$~$M_\odot$. Because most undiscovered members of the cluster 
are probably brown dwarfs, their inclusion would have negligible effect
on this estimate.
The combination of the mass of the stellar population with a mass of 
$\sim1000$~$M_\odot$ for the cloud \citep{bou98,miz01} implies a star 
formation efficiency of $\sim10$\% for Chamaeleon~I. This value is an upper
limit because the cloud was probably more massive in the past.

\section{Initial Mass Function}
\label{sec:imf}

\subsection{Defining the Sample}

In this section, I measure IMFs for the $1.5\arcdeg\times0.35\arcdeg$ and
$0.22\arcdeg\times0.28\arcdeg$ fields in Chamaeleon~I that are indicated
in Figure~\ref{fig:map1}.
Cluster members at higher masses can be detected through larger amounts of 
extinction. Therefore, to avoid IMF measurements that are biased in mass, 
I construct them from extinction-limited samples of known members. 
For the $1.5\arcdeg\times0.35\arcdeg$ and $0.22\arcdeg\times0.28\arcdeg$ 
fields, I select limits of $A_J\leq1.2$ and $A_J\leq1.4$, respectively. 
The current census of cluster members should be complete down to 0.03 
and 0.01~$M_\odot$ for these areas and ranges of extinctions 
(\S~\ref{sec:complete}).
The extinctions used in creating these extinction-limited
samples are those listed in Table~\ref{tab:hr} for the 215 members
with spectral classifications.
Eleven additional members lack extinction estimates because their spectral 
types are too uncertain, but they have very red near-IR colors that
are indicative of extinctions well beyond the limits adopted for the two IMFs. 
Pairs of members with separations less than $2\arcsec$ are treated as one 
source in the IMFs. In these cases, I adopt the masses of the earlier
components.
The six stars that appear below the main sequence in Figure~\ref{fig:hr} 
are rejected from the IMF samples because the census is not complete 
for subluminous objects of this kind.
After applying these criteria to the census of Chamaeleon~I, 
I arrive at IMF samples that contain 85 and 34 members for the
$1.5\arcdeg\times0.35\arcdeg$ and $0.22\arcdeg\times0.28\arcdeg$ fields, 
respectively.

\subsection{Estimating Masses}
\label{sec:masses}

For the objects in the IMF samples, I estimate masses from their 
positions on the H-R diagram by using the theoretical evolutionary models of 
\citet{bar98} and \citet{cha00} for $M/M_\odot\leq1$ and the models of 
\citet{pal99} for $M/M_\odot>1$ because they provide the best agreement with 
observational constraints \citep{luh03ic,luh05abdor,luh06}.
As noted in \S~\ref{sec:hr}, some of the late-type members of Chamaeleon~I 
have rather old isochronal ages in Figure~\ref{fig:hr}. 
For the purposes of this work, I assume that these old apparent ages are 
caused by errors in the luminosities, temperature scale, or models and
that these objects in reality have the same ages as the stellar members. 
Therefore, for members later than M7 that are below the 10~Myr isochrone, I 
estimate masses from their spectral types assuming that they have ages of 
1-3~Myr. 

\begin{figure}
\epsscale{.60}
\plotone{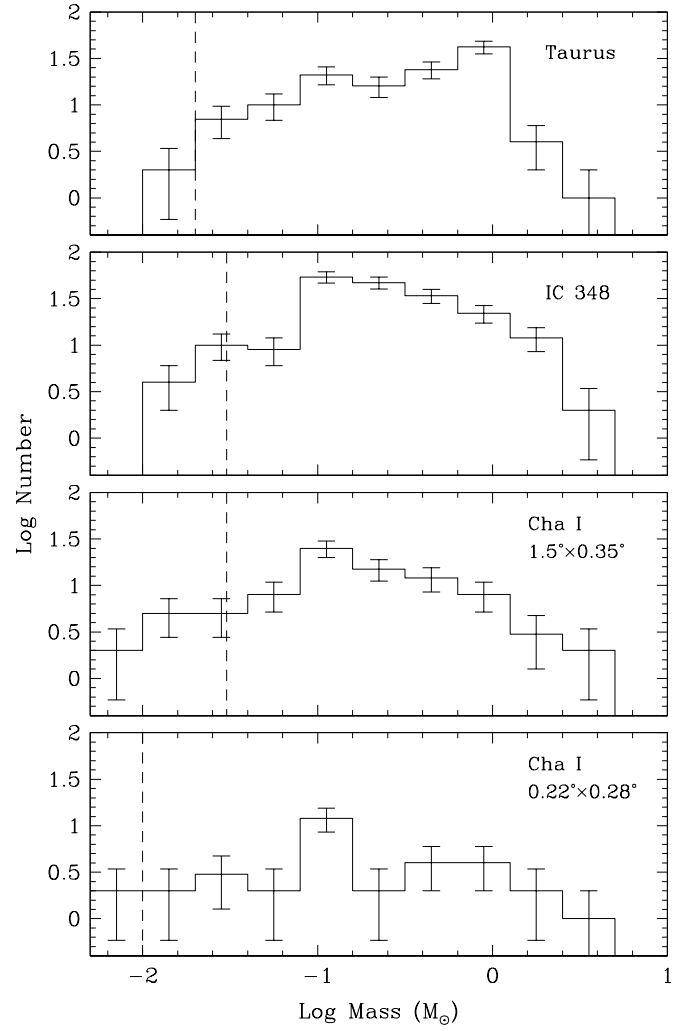}
\caption{
IMFs for extinction-limited samples in 
Taurus \citep{luh04tau}, IC~348 \citep{luh03ic}, and the 
$1.5\arcdeg\times0.35\arcdeg$ and $0.22\arcdeg\times0.28\arcdeg$ fields
in Chamaeleon~I in Figure~\ref{fig:map1}.
The completeness limits of these samples are indicated ({\it dashed lines}). 
In the units of this diagram, the Salpeter slope is 1.35.
}
\label{fig:imf}
\end{figure}

\begin{figure}
\plotone{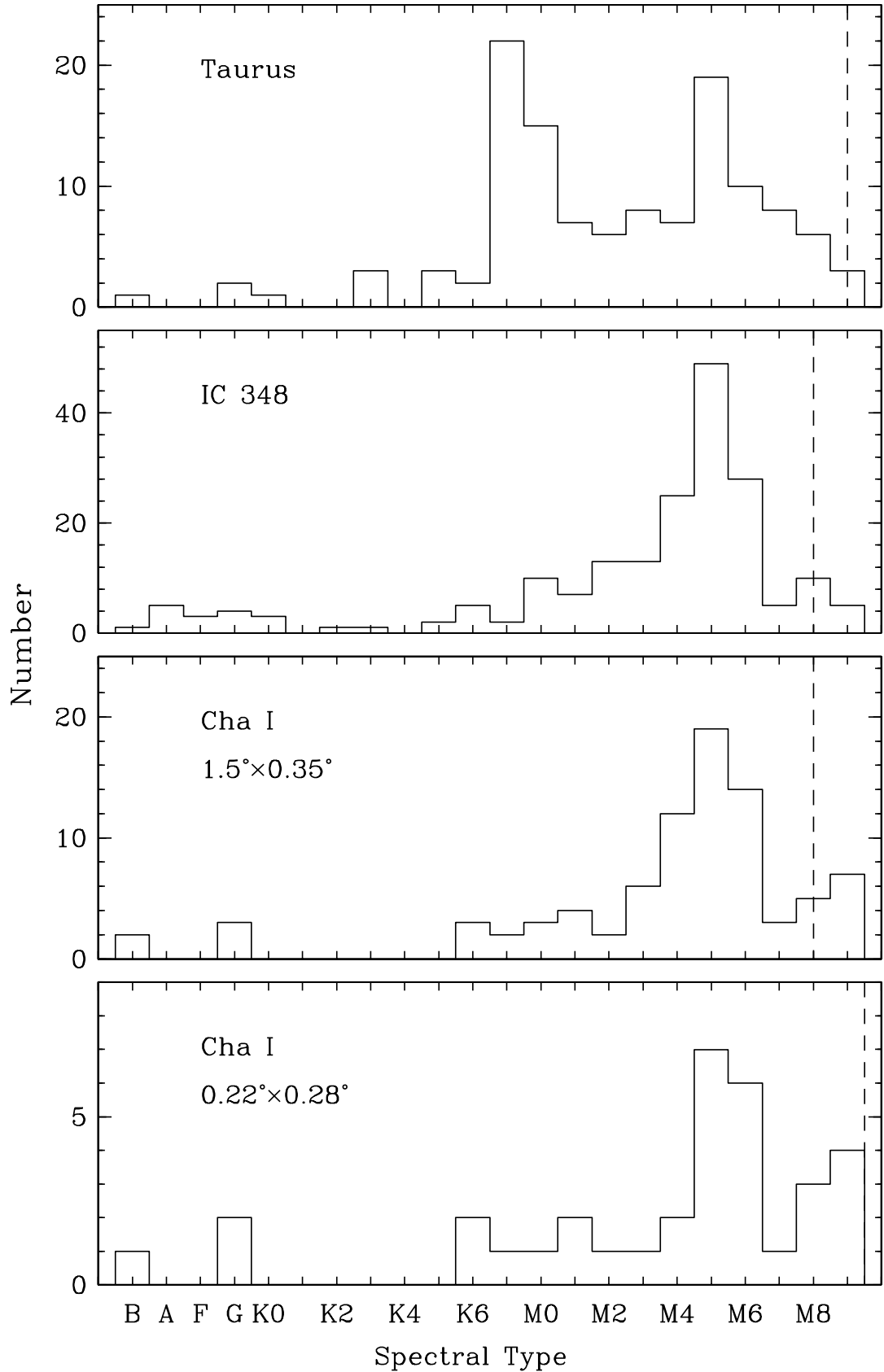}
\caption{
Distributions of spectral types for the IMFs in Figure~\ref{fig:imf}. 
The completeness limits of these samples are indicated ({\it dashed lines}). 
}
\label{fig:histo}
\end{figure}

\begin{figure}
\plotone{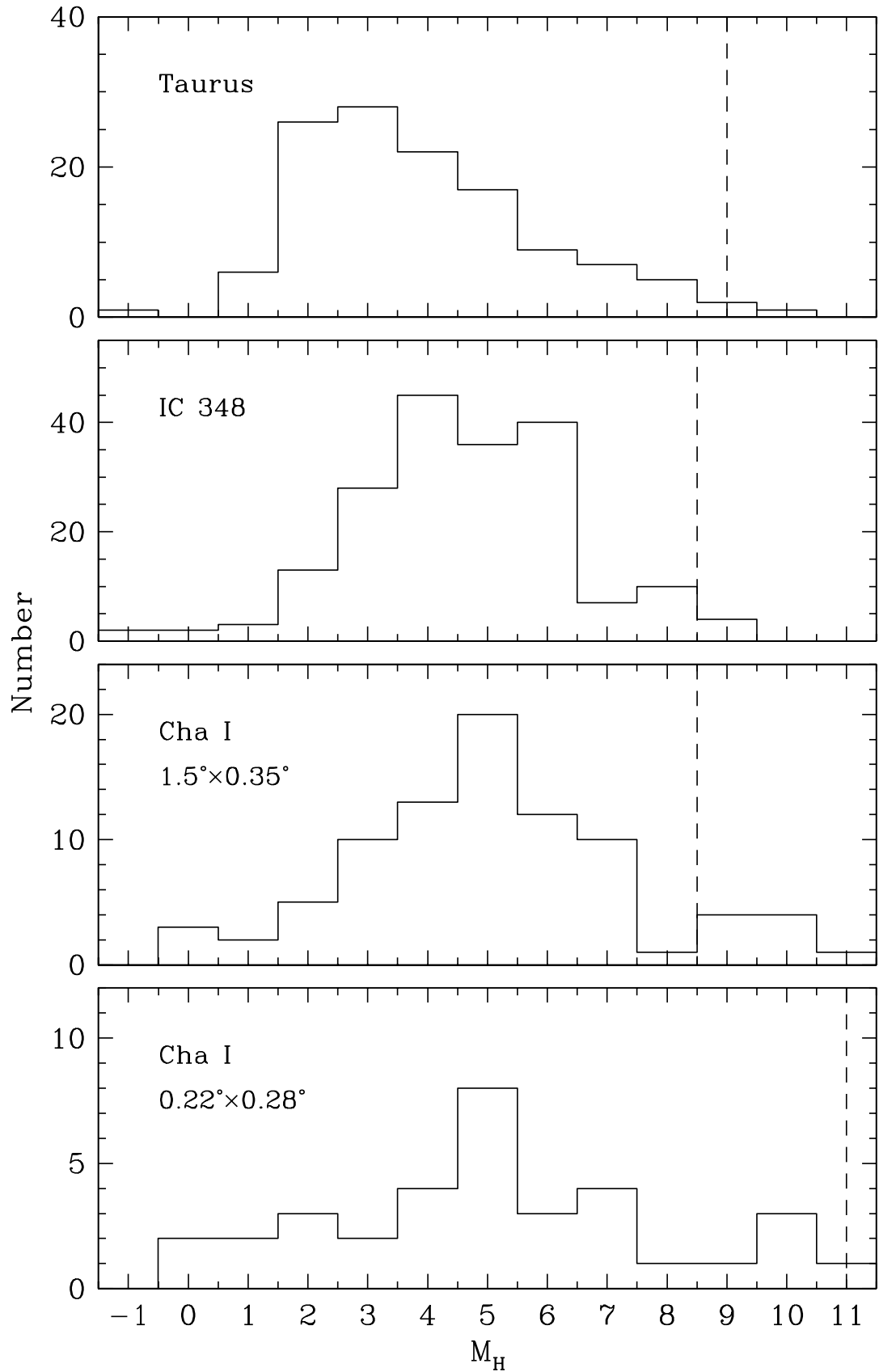}
\caption{
Distributions of $H$-band magnitudes for the IMFs in 
Figure~\ref{fig:imf}, corrected for extinction and distance.
The completeness limits of these samples are indicated ({\it dashed lines}). 
}
\label{fig:hlf}
\end{figure}

I estimate the masses of the known members classified as $\geq$M9
by comparing their luminosities to the values predicted for the evolutionary 
models of \citet{cha00} and \citet{bur97}. To account 
for the large range of isochronal ages exhibited by the late-type members
in Figure~\ref{fig:hr}, I adopt a conservatively large range of ages of 
1-30~Myr. Using these ages, the models imply masses of 
0.005-0.015~$M_{\odot}$ for the four $\geq$M9 members. 
In the IMFs, I divide these objects evenly between the two mass bins 
on either side of 0.01~$M_\odot$.
The resulting IMFs for the $1.5\arcdeg\times0.35\arcdeg$ and 
$0.22\arcdeg\times0.28\arcdeg$ fields are presented in Figure~\ref{fig:imf}.
For comparison, I include the similarly-derived IMFs for Taurus 
\citep{luh04tau} and IC~348 \citep{luh03ic}. For these IMF samples from 
Chamaeleon~I, Taurus, and IC~348, I also show the distributions of spectral 
types and dereddened absolute $H$-band magnitudes in Figures~\ref{fig:histo}
and \ref{fig:hlf}, respectively.

\subsection{Turnover Mass}

Notable features to examine in the IMF of a young cluster 
are the mass at which it reaches a maximum, the relative numbers 
of objects in different mass ranges, and the minimum mass. 
Because it contains more objects, the IMF in the $1.5\arcdeg\times0.35\arcdeg$ 
field provides a better measurement of the first two properties, while 
the deeper IMF in the $0.22\arcdeg\times0.28\arcdeg$ area better constrains 
the minimum mass of the IMF. 

The mass at which the IMF reaches a maximum has been
shown to vary significantly between the low-density stellar population
in Taurus \citep[$M\sim0.8$~$M_\odot$][]{luh00tau,luh04tau,bri02} and the
denser clusters of IC~348 in Perseus and the Orion Nebula Cluster
\citep[$M\sim0.1$-0.2~$M_\odot$][]{hil97,hc00,mue02,mue03,luh03ic}.
The peak mass of the $1.5\arcdeg\times0.35\arcdeg$ 
IMF in Chamaeleon~I is 0.1-0.15~$M_\odot$, which is similar to the value 
in IC~348, as shown in Figure~\ref{fig:imf}.
The agreement between Chamaeleon~I and IC~348 is also apparent in
the distributions of spectral types and M$_H$ in 
Figures~\ref{fig:histo} and \ref{fig:hlf}. Meanwhile, Taurus is distinct
from both clusters in all of these diagrams. 
Thus, the characteristic mass of the IMF exhibits no noticeable variation
among Orion, Chamaeleon, and Perseus. With a peak mass near 0.8~$M_\odot$, 
Taurus remains unique among well-studied nearby star-forming regions. 
As discussed by \citet{bri02} and \citet{luh03ic}, the distinctive 
nature of the IMF in Taurus may be a reflection of an unusually high average
Jeans mass. 

\subsection{Minimum Mass}

To date, measurements of IMFs in nearby star-forming regions have reached
mass limits of $\sim0.005$-0.02~$M_{\odot}$ through luminosity function 
modeling \citep{luh00trap,hc00,mue02,mue03,pre03,luc05} and spectroscopy
\citep{hil97,com00,bej01,luh03ic,bri02,bar04,luh04tau,sle04,lev06,luc06}.
Individual brown dwarfs that are not part of IMF measurements have also
been found near masses of $\sim0.01$~$M_{\odot}$ \citep{kir06,all06,all07}.
The best constraints on the minimum mass in Chamaeleon~I
are provided by my survey of a $0.22\arcdeg\times0.28\arcdeg$ field in the 
southern subcluster. The IMF for this field extends down to the completeness
limit of $\sim0.01$~$M_\odot$, and perhaps even to 0.005~$M_\odot$. 
Thus, the minimum mass of the IMF in Chamaeleon~I is $<0.01$~$M_\odot$,
which is comparable to constraints provided by IMF measurements in a few
other young clusters.
In comparison, a minimum mass of 0.001-0.01~$M_\odot$ has been predicted
by calculations of opacity-limited fragmentation 
\citep{low76,rees76,sil77,bos88,bat05,whi06}. To better constrain these
predictions with Chamaeleon~I, a survey with a deeper completeness limit
is required. 

\subsection{Brown Dwarf Fraction}

As done in \citet{bri02} and \citet{luh03ic}, I quantify the relative numbers 
of brown dwarfs and stars in Chamaeleon~I with the following ratio:

$$ {\mathcal R}_{1} = N(0.02\leq M/M_\odot\leq0.08)/N(0.08<M/M_\odot\leq10)$$

\noindent
For the $1.5\arcdeg\times0.35\arcdeg$ IMF, this ratio is
${\mathcal R}_{1} = 16/62 = 0.26^{+0.06}_{-0.05}$.
This brown dwarf fraction for Chamaeleon~I is higher than 
measurements in IC~348 and Taurus by a factor of 1.5-2
\citep{bri02,luh03tau,luh03ic,luh04tau} and is similar to 
values reported for Orion \citep{luh00trap,hc00,mue02,sle04}.
The variations among these ratios are significant if only the formal 
statistical errors are considered. However, it is likely that additional 
systematic errors are present in these measurements. In particular, 
because the peak of the IMF is close to the hydrogen burning limit 
(see Figs.~\ref{fig:imf} and \ref{fig:histo}), 
small systematic offsets in mass estimates can result in large differences 
in the relative numbers of stars and brown dwarfs. These offsets can arise from
differences in adopted evolutionary models, temperature scales, and spectral 
classification systems \citep{luh06tau}, or from basic differences in the
methods of estimating masses (e.g., positions on H-R diagrams versus luminosity 
functions). These sources of systematic errors are not present among
the measurements that I have performed in IC~348, Taurus, and Chamaeleon~I. 
However, because the low-mass stars in IC~348 and Chamaeleon~I are segregated
and low-mass stars dominate these stellar populations by number, the brown
dwarf fraction is also sensitive to the size of the field in which it is
measured. The fields considered in IC~348, Taurus, and Chamaeleon~I are 
probably large enough to accurately represent the entire stellar population in 
each cluster, but it is possible that the brown dwarf fractions would change 
noticeably if they were measured from even larger fields. 
Even in the absence of systematic errors, the modest differences in the 
brown dwarf fractions of these regions can be explained by
slight variations in the peak mass of the IMF and probably do not 
require a deeper explanation that is specific to brown dwarf formation.

\section{Spatial Distribution}
\label{sec:spatial}

\subsection{Brown Dwarfs within $3\arcdeg$}

If I account for the completeness limits described in 
\S~\ref{sec:complete}, I can use the current census of Chamaeleon~I to 
characterize the spatial distribution of stars and brown dwarfs in this cluster.
I first investigate whether a widely-distributed population of brown dwarfs
is present in Chamaeleon~I.
The IK2 survey was specifically designed for this question (\S~\ref{sec:ik2}). 
To be sensitive to members moving at high 
velocities ($v\gtrsim2$~km~s$^{-1}$), such as brown dwarfs ejected from 
multiple systems \citep{rc01}, I selected a search radius of $3\arcdeg$ from 
the center of Chamaeleon~I, which corresponds to the angular distance traveled 
by an object moving at 8~km~s$^{-1}$ in the plane of the sky at the distance of 
Chamaeleon~I for a duration of 1~Myr. I also considered only sources with 
$i-K_s>2.5$ ($\gtrsim$M4.5), which is blue enough to encompass the peak of 
the spectral type distribution of the cluster (Figure~\ref{fig:histo}) while 
red enough to minimize contamination by field stars (Figure~\ref{fig:ik2}). 
As demonstrated in \S~\ref{sec:complete}, the IK2 survey is complete for 
substellar cluster members above masses of 0.03~$M_\odot$ ($\lesssim$M8) that 
are within areas covered by the Second DENIS Release (Figure~\ref{fig:map1}).

As shown in the map of known members of Chamaeleon~I in Figure~\ref{fig:map1}b,
a population of widely-distributed brown dwarfs is not present
out to a radius of $3\arcdeg$ (8.5~pc) surrounding the cluster. 
To quantify this result, I compare the relative numbers of low-mass stars 
and brown dwarfs between
the inner and outer regions of the cluster. The ratio N($>$M6-M8)/N(M4.5-M6)
is $1/12=0.08^{+0.15}_{-0.02}$ for $1.25<r\leq3\arcdeg$, which is actually 
lower than the value of $15/70=0.21^{+0.06}_{-0.05}$ for $r\leq1.25\arcdeg$.

Several studies have suggested that brown dwarfs form by
the ejection of protostellar sources from multiple systems
\citep{rc01,bos01,bat02,del03,umb05,goo07}. Some of these models predict that
brown dwarfs are born with higher velocity dispersions than stars, and
consequently have wide spatial distributions in star-forming regions
\citep{rc01,kb03}. However, I have shown that brown dwarfs are not more widely 
distributed than their stellar counterparts in Chamaeleon~I.
\citet{luh06tau} and \citet{sle06} found the same result for Taurus, 
which also lacks an extended population of brown dwarfs. Thus, embryo 
ejection models that predict high velocities for brown dwarfs are not viable
as the dominant mode of brown dwarf formation.

\subsection{Stars and Brown Dwarfs within $1\arcdeg$}

I next examine the spatial distributions of members of Chamaeleon~I within
a radius of $\sim1\arcdeg$ from the cluster center.
Because I wish to compare the distributions for different ranges of spectral
types, I begin by summarizing the completeness within this area
(\S~\ref{sec:complete}). 
The current census of the central $1.5\arcdeg\times0.35\arcdeg$ field 
should be nearly 100\% complete for $M/M_{\odot}\geq0.03$ ($\lesssim$M8)
and $A_J\leq1.2$. Outside of this field, most members at M4.5-M8 should be
identified. Members later than M8 are excluded from this analysis of 
spatial distributions because only the central regions of the two
subclusters in Chamaeleon~I have been thoroughly searched at such late types. 
The completeness at types earlier than mid-M is not well-determined
outside of the $1.5\arcdeg\times0.35\arcdeg$ field, but 
it is likely that wide-field X-ray and H$\alpha$ surveys have 
found most of the earlier members, particularly given the low levels of 
extinction outside of the $1.5\arcdeg\times0.35\arcdeg$ field. 
Thus, spatial distributions measured from members earlier than M8 in the
current census should be accurate representations of the cluster.

\begin{figure}
\epsscale{1.0}
\plotone{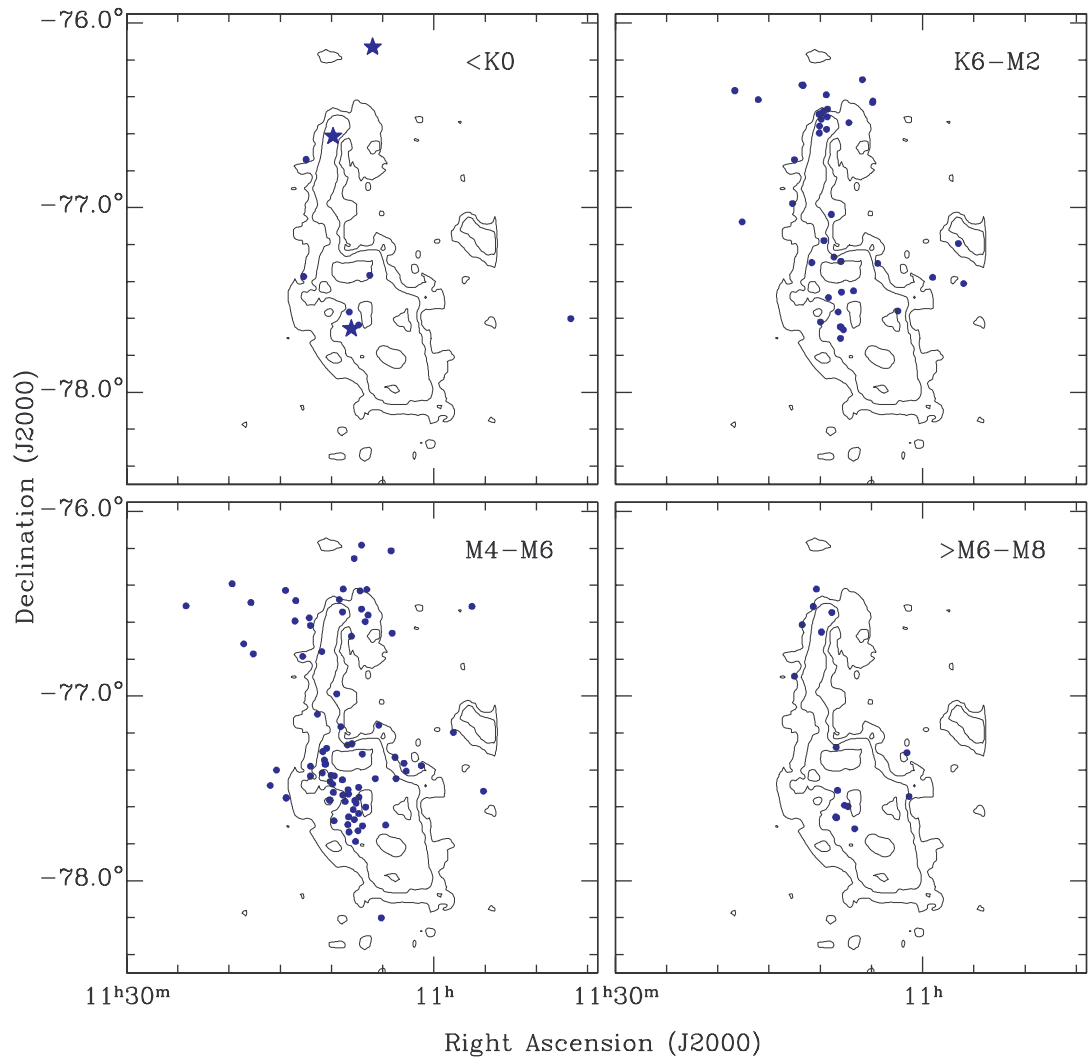}
\caption{
Spatial distributions of members of Chamaeleon~I for different ranges
of spectral types. The three B stars are indicated ({\it stars}).
The contours represent the extinction map of \citet{cam97} at intervals
of $A_J=0.5$, 1, and 2.
}
\label{fig:map5}
\end{figure}

To characterize the spatial distribution of the stellar population
of Chamaeleon~I as a function of mass, I first plot the positions of members 
with spectral types of $<$K0, K6-M2, M4-M6, and $>$M6-M8 on separate maps
in Figure~\ref{fig:map5}.
The cluster contains too few stars at $<$K0 to examine their spatial 
distribution in detail. I only note that two of the three most massive
members, the B stars T32 and T41, are located near the centers of the two
subclusters, which is often the case in young clusters 
\citep[][references therein]{hil98}. The third B star, HD~96675, is north
of the Chamaeleon~I cloud (see Figure~\ref{fig:map5}). 
In \S~\ref{sec:morph}, I suggested that the
it may have formed from small cloudlet that has recently dissipated.

\begin{figure}
\plotone{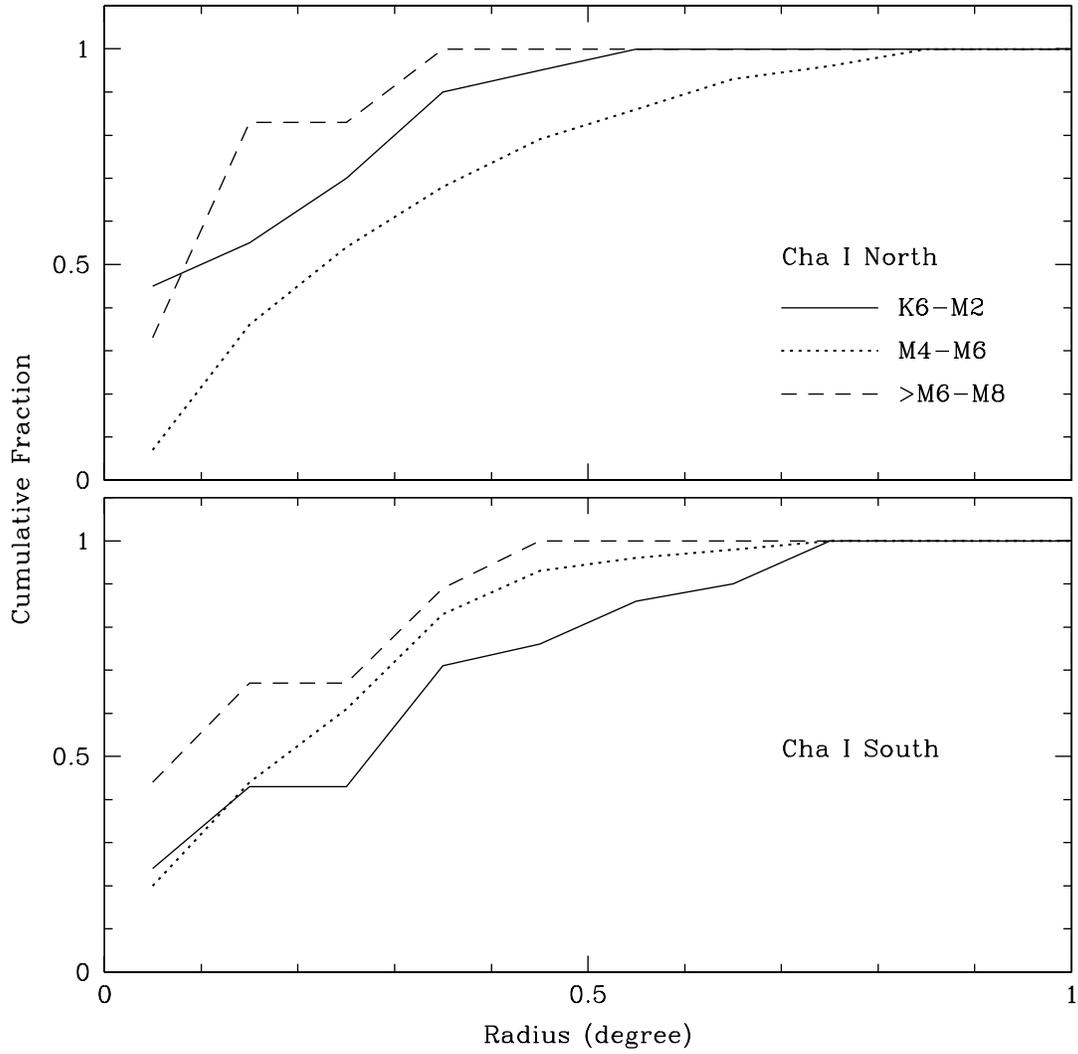}
\caption{
Cumulative distributions of projected angular radii from the centers of the 
northern ({\it top}, $\delta>-77\arcdeg$) and southern ({\it bottom},
$\delta<-77\arcdeg$) subclusters in Chamaeleon~I for members with spectral
types of K6-M2 ({\it solid lines}), M4-M6 ({\it dotted lines}), and 
$>$M6-M8 ({\it dashed lines}). 
The centers are defined as the positions of the maximum surface densities,
corresponding to 
$\alpha=11^{\rm h}09^{\rm m}40^{\rm s}$, $\delta=-76\arcdeg33\arcmin00\arcsec$
and 
$\alpha=11^{\rm h}08^{\rm m}00^{\rm s}$, $\delta=-77\arcdeg38\arcmin00\arcsec$
(J2000) for the northern and southern subclusters, respectively. 
An outer radius of $1\arcdeg$ was adopted for each subcluster.
}
\label{fig:rad}
\end{figure}

Based on visual inspection of Figure~\ref{fig:map5}, the members at K6-M2,
M4-M6, and $>$M6-M8 exhibit similar spatial distributions in the
southern subcluster. However, in the northern subcluster, the M4-M6 members 
appear to have a significantly wider distribution than the objects at K6-M2 
and $>$M6-M8. To quantitatively assess this possible mass segregation, I have
computed the cumulative distributions of radii from the centers of the
northern and southern subclusters for K6-M2, M4-M6, and $>$M6-M8, which
are shown in Figure~\ref{fig:rad}. Again, the northern M4-M6 members
are less centrally concentrated than the other populations.
According to two-sided Kolmogorov-Smirnov tests among the northern radial 
distributions, the probability that the M4-M6 sample is drawn from 
the same parent distribution as K6-M2 and $>$M6-M8 is 1-2\%, while the
latter two samples are consistent with each other. Meanwhile, no significant
differences are present among the three radial distributions in the southern 
subcluster.

Mass segregation of the kind found in the northern subcluster
has been reported in other young clusters such as IC~348 \citep{mue03} and the 
Orion Nebula Cluster \citep{hc00}. In each of those clusters, the peak mass of 
the IMF varies from $\sim0.5$~$M_\odot$ in the core to $\sim0.1$~$M_\odot$ at 
large radii. Dynamical cluster evolution is not a plausible source of the 
mass segregation of low-mass stars in IC~348 and Orion, and it is even
less likely at the lower stellar densities found in Chamaeleon~I.
Instead, it would seem that the mass segregation in these clusters must 
be primordial. However, it is unclear why the southern subcluster shows
no segregation. Measurements of velocities for the members of Chamaeleon~I
and an analysis of their dependence on mass and position would likely prove to
be valuable in further exploring this issue.

\subsection{Constraints on Star Formation History}

The spatial distribution of members of Chamaeleon~I can also be used
to constrain the star formation history.
To search for a widely distributed population of members,
\citet{cov97} measured Li absorption strengths and radial
velocities for X-ray sources from the {\it ROSAT} all-sky survey
within 170 square degrees surrounding the Chamaeleon complex of clouds. 
They identified 37 pre-main-sequence stars, which exhibited spectral types of 
late G to early M. Complimenting that study, my wide-field survey 
(IK2, \S~\ref{sec:ik2}) considers a smaller area but reaches lower 
masses, and most importantly encompasses the peak of the IMF. 
These two surveys indicate that few young stars reside at large 
distances from the Chamaeleon~I cloud.

\citet{fei96} suggested that star formation has occurred at a constant
rate for 20~Myr in Chamaeleon~I and that the observed deficit of older stars 
($\tau\gtrsim2$~Myr) in this cluster and other star-forming regions 
is caused by the dispersal of the older stars. 
However, in contrast to the expectations of that scenario, the wide-field
surveys with {\it ROSAT} \citep{alc95,alc97,cov97} and in this work have 
shown that the Chamaeleon~I clouds are not surrounded by a large population
of young stars. Similar results have been found for other star-forming regions
\citep{bri97}. In addition, the distributions of isochronal ages in
Figure~\ref{fig:ages} are inconsistent with a constant star formation
rate for 10-20~Myr.

\section{Conclusions}

I have presented an extensive search for new members of the Chamaeleon~I 
star-forming region. 
Because the completeness limits of my survey are well-determined, I have 
been able to perform robust measurements of the distributions of members 
of Chamaeleon~I as a function of mass, position, and age. 
The primary results of this study are summarized as follows:

\begin{enumerate}

\item
I have discovered 50 new members of Chamaeleon~I, which increases the
census of known members to 226 objects. The new members include
14 objects that are later than M6 ($M\lesssim0.08$~$M_\odot$) and the
two faintest known members of the cluster, which may have masses of 
only 0.005-0.01~$M_\odot$. The current census now contains 28 members 
that are likely to be substellar. 

\item
The distribution of isochronal ages for members of Chamaeleon~I between
0.1-1~$M_\odot$ suggests that star formation has occurred for the past
3-4 and 5-6~Myr in the southern and northern subclusters, respectively, 
at rates that have declined with time.

\item
The IMF in Chamaeleon~I reaches a maximum at a mass of 0.1-0.15~$M_\odot$, 
which is similar to the turnover mass observed in IC~348 and the Orion Nebula 
Cluster \citep{hil97,hc00,mue02,mue03,luh03ic}. 
The substellar IMF is roughly flat in logarithmic units and shows no indication 
of reaching a minimum down to a completeness limit of 0.01~$M_\odot$. 

\item
Chamaeleon~I does not contain a widely-distributed population of brown dwarfs,
which is contrary to the predictions of some embryo ejection models. 
Instead, the substellar
members share the same spatial distribution as the stars. However, low-mass
stars in the northern subcluster do appear to have a wider distribution
than members at other masses, which resembles the mass segregation that has
been previously observed in Orion and IC~348 \citep{hc00,mue03}.

\end{enumerate}

\acknowledgements
I thank the staff at Las Campanas Observatory for their assistance with these 
observations and Eric Feigelson and Eric Mamajek for helpful comments on
the manuscript. This work was supported
by grant NAG5-11627 from the NASA Long-Term Space Astrophysics program and 
grant GO-10138 from the Space Telescope Science Institute. 
This publication makes use of data products from 2MASS, which
is a joint project of the University of Massachusetts
and the Infrared Processing and Analysis Center/California Institute
of Technology, funded by NASA and the NSF.
The DENIS project has been partly funded by the SCIENCE and the HCM plans of
the European Commission under grants CT920791 and CT940627.
It is supported by INSU, MEN and CNRS in France, by the State of 
Baden-W\"urttemberg in Germany, by DGICYT in Spain, by CNR in Italy, by 
FFwFBWF in Austria, by FAPESP in Brazil, by OTKA grants F-4239 and F-013990 
in Hungary, and by the ESO C\&EE grant A-04-046.
Jean Claude Renault from IAP was the Project manager. Observations were  
carried out thanks to the contribution of numerous students and young 
scientists from all involved institutes, under the supervision of P. Fouqu\'e.

\appendix

\section{Comments on Individual Sources}
\label{sec:app1}

The strengths of the gravity-sensitive lines in the spectra of
2MASS~10555824-7418347 and 10435748-7633023 appear to be intermediate
between those of field dwarfs and known Chamaeleon~I members, indicating
that these objects may be young members of the field.
Similar objects have been identified toward the Taurus cloud complex 
\citep{luh06tau,sle06} and in the solar neighborhood \citep{cru07}.

\begin{figure}
\epsscale{.80}
\plotone{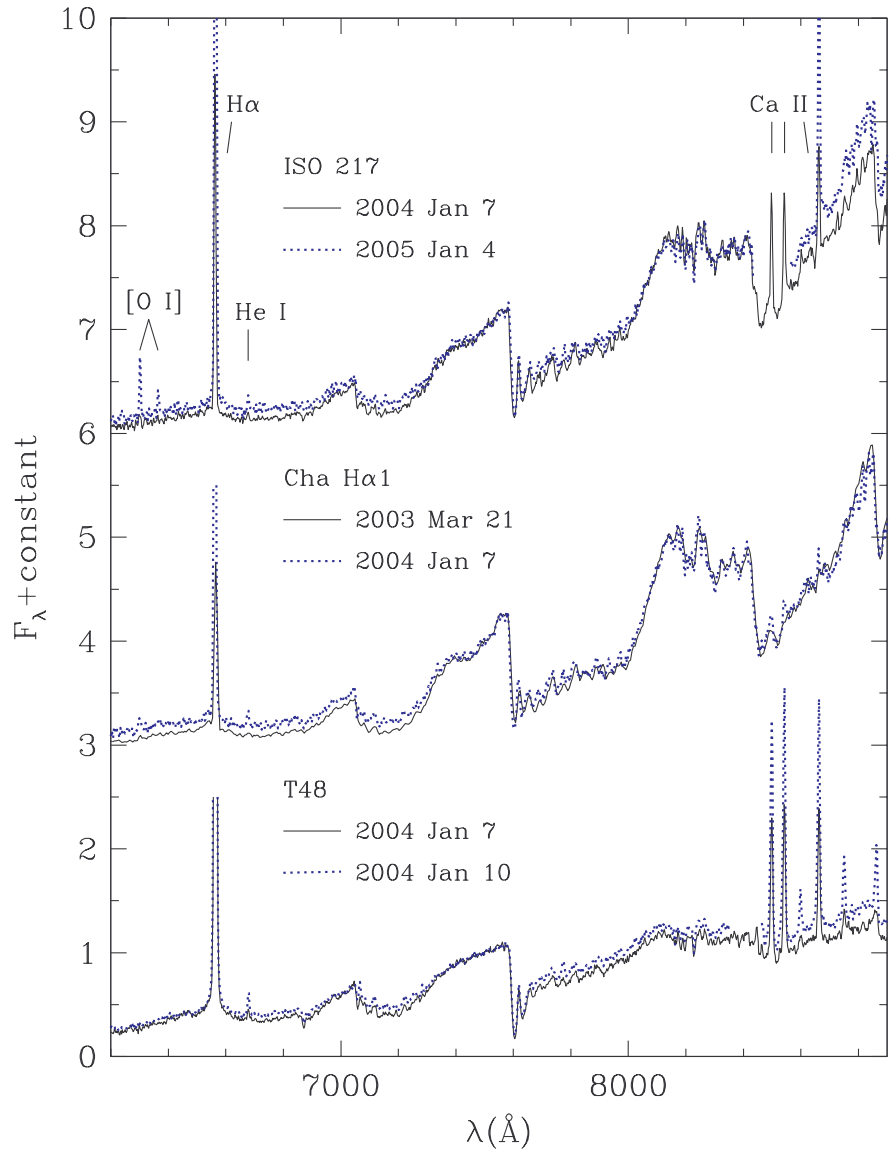}
\caption{
Chamaeleon~I members that exhibit variability in their spectra. 
For each object, an increase in H$\alpha$ and other emission lines is 
accompanied by brighter continuum emission and weaker absorption bands
at blue wavelengths, which are indicative of veiling by blue excess emission.
The data are normalized at 7500~\AA.}
\label{fig:op7}
\end{figure}

The spectra of ISO~217 from \citet{luh04cha} and from January 2004 are similar, 
but the spectrum from January 2005 for this object exhibits stronger emission 
lines, brighter blue continuum, and weaker TiO than those two earlier spectra.
All of these characteristics are consistent with an increase in the accretion
rate. Similar variability is present in the data for Cha~H$\alpha$~1 and T48,
as shown in Figure~\ref{fig:op7}. The variability for T48 occurred over a
period of only three days.

ESO~H$\alpha$~281 was resolved as a $1.7\arcsec$ pair by \citet{rz93}.
The brighter component was designated as ESO~H$\alpha$~281~A by \citet{bz97}
and is 2MASS~11070350-7631443.
The spectral classifications from \citet{bz97} and in this work indicate
that this star is a background giant. 
The fainter component, ESO~H$\alpha$~281~B from \citet{bz97}, is not resolved
from the brighter star by 2MASS. 
It was spectroscopically confirmed as a member of Chamaeleon~I by \citet{bz97}, 
but it was overlooked in the compilation of members in \citet{luh04cha}.

The spectrum of 2MASS~11253653-7700348 has M-type spectral features but
does not match any standard dwarfs, giants, or pre-main-sequence objects. 
Thus, the luminosity class of this object is uncertain. It is unlikely to be
a young star based on the absence of Li absorption in its spectrum.

The strengths of the Na~I and K~I lines in the spectrum of 
2MASS~11052472-7626209 are suggestive of youth. However, because 
the variation of these lines between dwarfs and pre-main-sequence stars
is subtle for the spectral type of this star (M2.75), it remains 
possible that it is a field dwarf. Similarly, 
the triangular $H$-band continuum of Cha~J11083040-7731387 is 
indicative of youth \citep{luc01},
but this object appears to have strong K~I absorption
at 1.25~\micron, which suggests higher, dwarf-like gravity.
Because no spectral type of field M and L dwarfs matches the spectrum of
Cha~J11083040-7731387 and the apparent strength of K~I may be due to 
low signal-to-noise, I assume that it is a member of Chamaeleon~I for the
purposes of this work.

\begin{figure}
\plotone{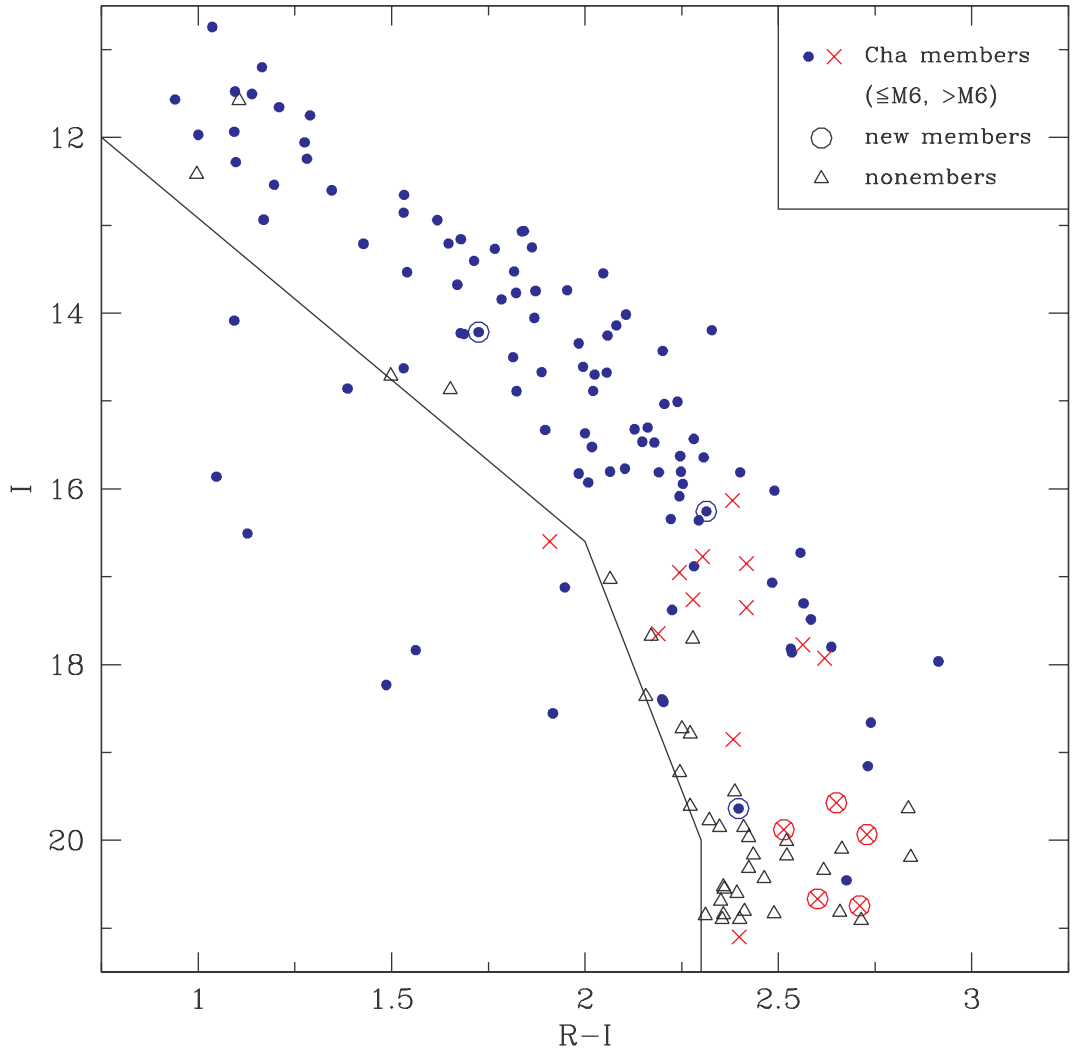}
\caption{
Color-magnitude diagram for the RI survey of Chamaeleon~I, which applies 
to the fields imaged with IMACS in Figure~\ref{fig:map1}c.
Candidate members of Chamaeleon~I were selected based on positions above and
right of the solid boundary and were spectroscopically classified as new members
({\it circles}) or nonmembers ({\it triangles}).
All known members within this field are indicated 
({\it filled circles and crosses}).
}
\label{fig:ri}
\end{figure}

\begin{figure}
\epsscale{1.0}
\plotone{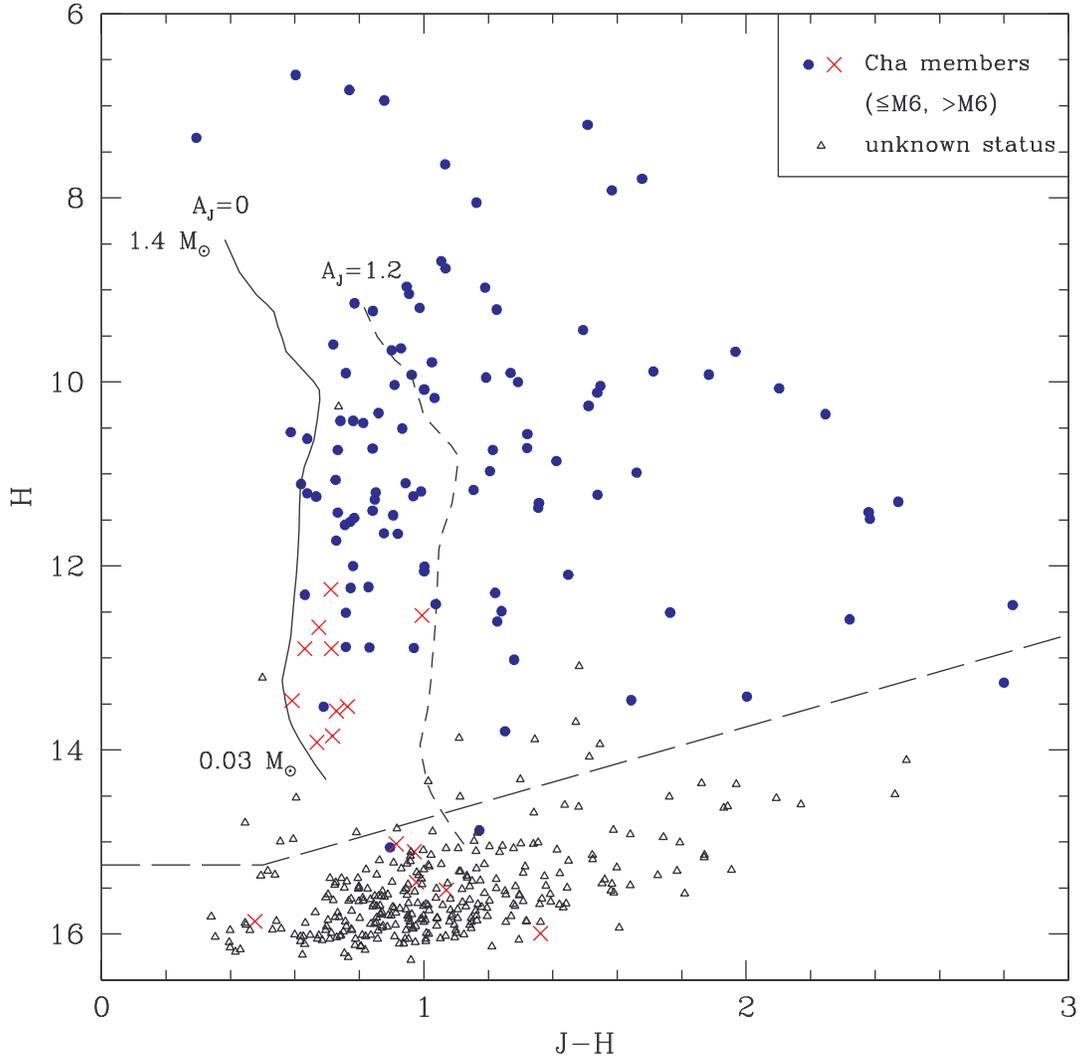}
\caption{
$J-H$ versus $H$ from 2MASS for the $1.5\arcdeg\times0.35\arcdeg$ field 
in Figure~\ref{fig:map1}a.
I have omitted the stars classified as nonmembers through spectroscopy
(Tables~\ref{tab:non}) as well as objects that are probable field stars
because they are below the sequences of known members 
in Figures~\ref{fig:ik1} and \ref{fig:ri}.
This diagram shows the remaining stars, which consist of 
known members of Chamaeleon~I  ({\it filled circles and crosses})
and stars whose membership is undetermined ({\it triangles}). 
The 10~Myr isochrone (1.4-0.03~$M_{\odot}$) from the evolutionary models of
\citet{bar98} is shown for $A_J=0$ ({\it solid line}) and $A_J=1.2$ ({\it short
dashed line}). 
These measurements have completeness limits of $J=15.75$ and $H=15.25$
({\it long dashed line}).
}
\label{fig:jh1}
\end{figure}

\begin{figure}
\plotone{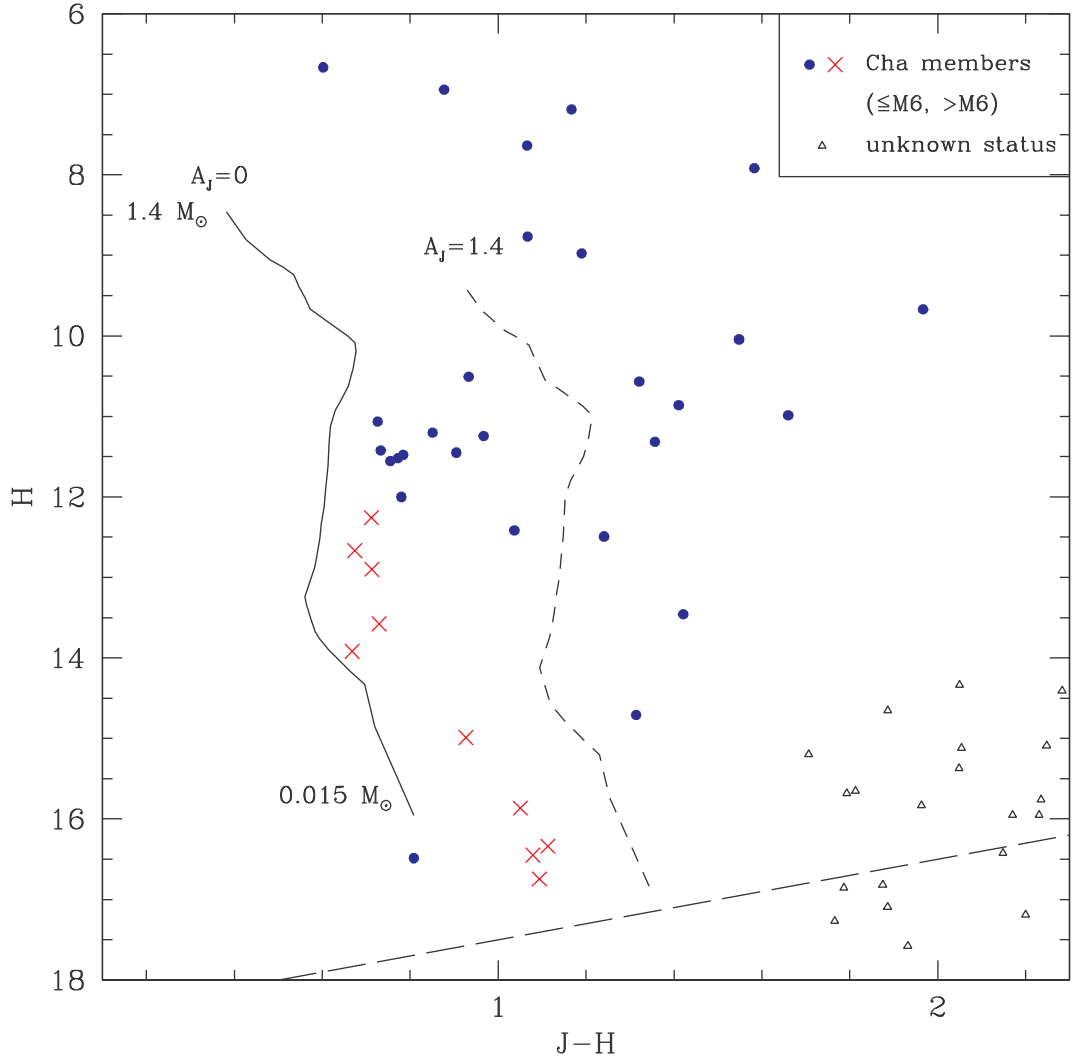}
\caption{
$J-H$ versus $H$ from 2MASS and ISPI for the $0.22\arcdeg\times0.28\arcdeg$ 
field imaged with ACS in Figure~\ref{fig:map1}d.
I have omitted the stars classified as nonmembers through spectroscopy
(Tables~\ref{tab:non}), objects that are probable field stars
because they are below the sequences of known members
in Figures~\ref{fig:ik1}, \ref{fig:ri}, and \ref{fig:iz}, and sources that are
resolved as galaxies in the ACS images.
This diagram shows the remaining stars, which consist of 
known members of Chamaeleon~I ({\it filled circles and crosses})
and stars whose membership is undetermined ({\it triangles}). 
The 10~Myr isochrone (1.4-0.015~$M_{\odot}$) from the evolutionary models of
\citet{bar98} is shown for $A_J=0$ ({\it solid line}) and $A_J=1.4$ ({\it short
dashed line}). 
The measurements have completeness limits of $J=18.5$ and $H=18.25$
({\it long dashed line}).
}
\label{fig:jh2}
\end{figure}

\citet{saf03} presented an optical spectrum for a putative new member
of Chamaeleon~I. They attributed their spectrum to 2MASS~11095300-7730588
and named the object SGR1. Their spectrum exhibited strong H$\alpha$ 
emission and they reported a spectral type of M7 from those data. 
However, my spectrum of this star shows no H$\alpha$ emission
or late-type molecular bands, and instead is indicative of a
reddened background field star. If placed on the color-magnitude diagram 
in Figure~\ref{fig:ri} ($R-I=1.4$, $I=18.7$), this star falls well below
the sequence of known members, which further indicates that it is a field
star rather than a member.  Finally, SGR1 should have a circumstellar 
disk if it has the strong H$\alpha$ emission found by \citet{saf03}.
However, SGR1 does not exhibit the excess emission expected from a disk
in mid-IR images from \citet{luh05frac}. The spectrum published by \citet{saf03}
(from multi-object spectroscopy) was probably incorrectly associated with
2MASS~11095300-7730588, and in fact may apply to a previously known member.

\begin{figure}
\epsscale{1.0}
\plotone{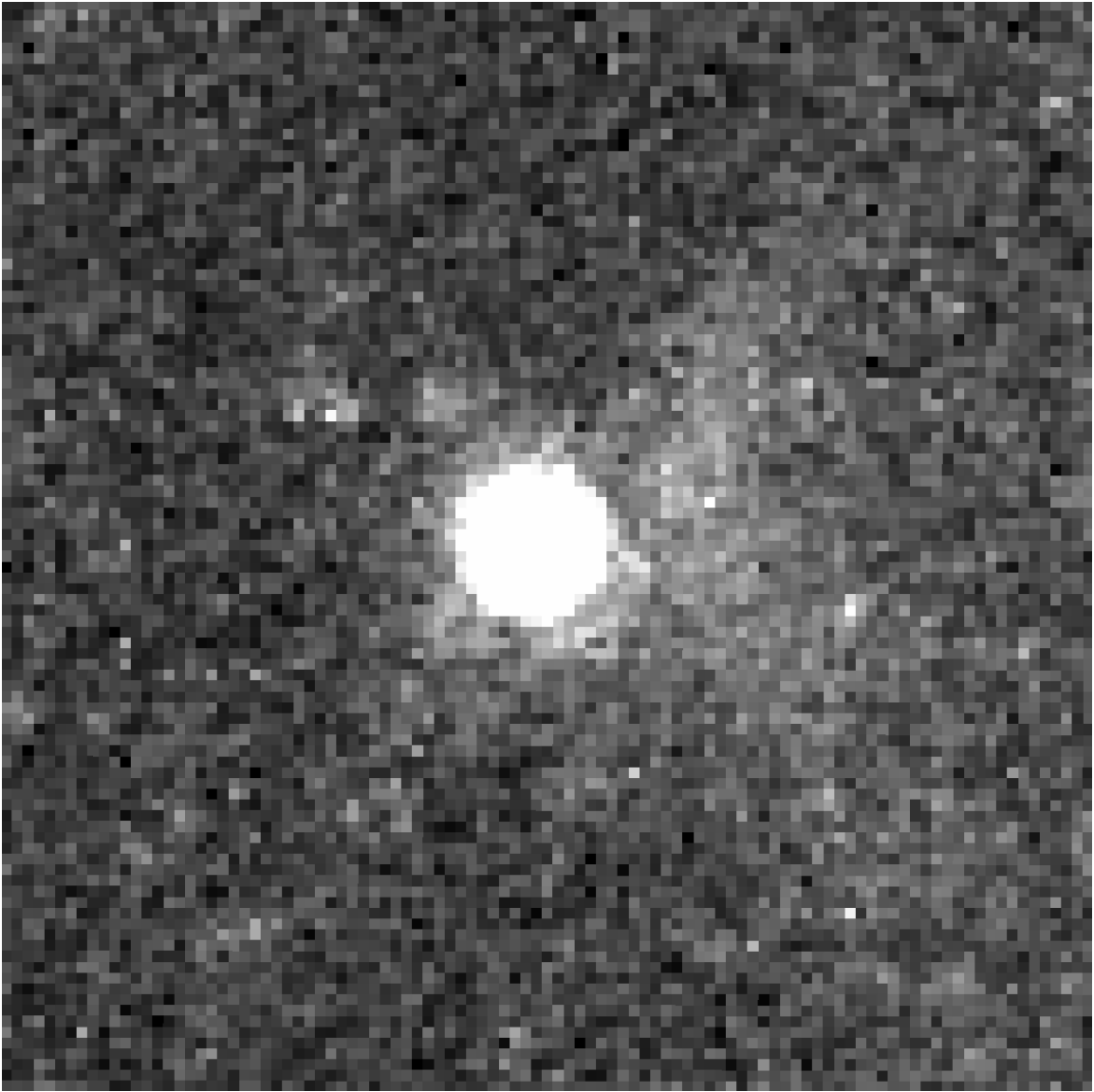}
\caption{
ACS F775W image of Cha~J11081938-7731522. The presence of extended emission 
and its butterfly morphology indicate that this star is seen in scattered
light (e.g., edge-on disk), which is consistent with its subluminous nature
(Figure~\ref{fig:iz} and \ref{fig:jh2}). 
The size of the image is $5\arcsec\times5\arcsec$.}
\label{fig:72422}
\end{figure}

Six known members of Chamaeleon~I appear below the main sequence in the
H-R diagram in Figure~\ref{fig:hr}, consisting of CHSM~15991, T14A, ISO~225,
ESO~H$\alpha$~569 and 574, and Cha~J11081938-7731522.
Low luminosity estimates were also derived by \citet{luh04cha}
for the first three stars and by \citet{com04} for ESO~H$\alpha$~569 and 574.
\citet{luh04cha} suggested that CHSM~15991, T14A, and ISO~225 are unusually
faint because they are seen only in scattered light (e.g., edge-on disks),
and \citet{com04} offered the same explanation for ESO~H$\alpha$~574. 
\citet{com04} concluded that ESO~H$\alpha$~569 is more likely to
have an intrinsically low luminosity rather than an edge-on disk
because excess emission from a disk was not detected in mid-IR observations 
with ISO \citep{per00}. However, ESO~H$\alpha$~569 
does exhibit strong excess emission in the {\it Spitzer} data from 
\citet{luh05frac}.  The absence of a detection of this star
in the X-ray images from \citet{fl04} can be used to constrain the amount
of extinction toward the star. 
If ESO~H$\alpha$~569 has an X-ray luminosity in the range of values observed
for classical T Tauri stars at masses of 0.4~$M_\odot$ \citep{tel07}, then 
the detection limit described by \citet{fl04} indicates 
an extinction of $A_K\gtrsim60$ toward this star.
Thus, the X-ray data from \citet{fl04} support the presence of an edge-on
disk around ESO~H$\alpha$~569.
The sixth star below the main sequence in Figure~\ref{fig:hr}, 
Cha~J11081938-7731522, is newly identified as a cluster member in this work.
It also appears subluminous in Figures~\ref{fig:ri}, \ref{fig:iz}, and
\ref{fig:jh2}. Although it falls below the sequences of known members in
those diagrams, Cha~J11081938-7731522 was not rejected as a field star 
because it was red enough in Figure~\ref{fig:ri} to be selected as a 
candidate (substellar) member. The presence of extended emission around 
this star in the ACS image in Figure~\ref{fig:72422} supports the idea
that it is seen in scattered light. In particular, the shape of the
emission resembles the butterfly morphology that characterizes edge-on disks.

\citet{kra07} presented a tabulation of 14 candidate wide companions to 
known members of Chamaeleon~I. My photometry and spectroscopy provide
new constraints on the cluster membership of some of these candidates. 
Through spectroscopy, I have classified the candidate brown dwarf companion
to ISO~250, 2MASS~11103749-7722083, as a field star. Note that this candidate
is different from the candidate companion to ISO~250 that was found by
\citet{lm04}. I also have spectroscopically classified the candidate
near KG~102 as a field star. 
The candidate companions to C7-1 and T14A are below the
cluster sequence in Figure~\ref{fig:ri} ($I=18.1$ and 18.4, $R-I=1.8$ and 0.8), 
indicating that they are probably not members of the cluster.
The candidate companions to T3 and T26 are spectroscopically confirmed as
members in this work. The candidate companion to T39A+B, 2MASS~11091297-7729115,
is a candidate member according to my color-magnitude diagrams. It is the
one remaining bright candidate appearing in Figure~\ref{fig:jh1}.
Although it was included in the list of candidate companions from
\citet{kra07}, the candidate near T33 was already confirmed as a cluster
member by previous studies \citep[e.g.,][]{luh04cha}.

\section{Comparison to Previous Surveys}
\label{sec:app2}

\subsection{\citet{fl04} and \citet{stel04}}
\label{sec:fl04}

Using the {\it Chandra X-ray Observatory}, \citet{fl04} obtained an
image of a $16\arcmin\times16\arcmin$ field in the northern subcluster of 
Chamaeleon~I, which is indicated in Figure~\ref{fig:map1}a.
They detected 27 previously known members and identified no new candidate 
members. \citet{fl04} concluded that the census of known members in their 
survey field is complete to masses of $\sim$0.1~$M_\odot$.
The membership classifications based on the X-ray data from \citet{fl04} agreed 
well with the classifications by \citet{luh04cha} using other diagnostics.
ISO~165 was the only star that was classified differently between the two 
studies. \citet{luh04cha} classified ISO~165 as a member of Chamaeleon~I
because of its strong H$\alpha$ emission ($W_{\lambda}=49\pm$3) and
weak Na~I and K~I lines that are indicative of low surface gravity.
In addition, it cannot be a foreground dwarf because it significantly reddened, 
while it cannot be a background dwarf because such an object would appear 
below the main sequence when placed on the H-R diagram at the distance of
Chamaeleon~I, which is not the case. 
In comparison, \citet{fl04} suggested that ISO~165 might be a nonmember
because it was not detected in their X-ray data. 
However, the spectral type of M5.5 for this object corresponds to a mass
of 0.1-0.15~$M_\odot$, which is near their completeness limit, so it is
not surprising that it was not detected. In the mid-IR images from
\citet{luh05frac}, ISO~165 exhibits mid-IR excess emission that indicates
the presence of a disk, which provides further evidence of the youth
and membership of this object in Chamaeleon~I. 
In addition to ISO~165, seven other known members of Chamaeleon~I were
not detected in the X-ray data from \citet{fl04}. Six of these 
sources have spectral types later than M7 ($M\lesssim0.06$~$M_\odot$), 
and thus detections were not expected since they fall below the 
mass completeness limit of the X-ray data. The seventh member, 
ESO~H$\alpha$~569 \citep{com04}, has a relatively high mass of 
0.4~$M_\odot$ based on its spectral type of M2.5, and yet it was
not detected by \citet{fl04}, which suggests the presence of extremely 
high extinction toward this star ($A_K\gtrsim60$, \S~\ref{sec:app1}).

Using {\it XMM-Newton}, \citet{stel04} obtained an X-ray 
image of a $27\arcmin\times27\arcmin$ field in the southern subcluster of 
Chamaeleon~I, which is indicated in Figure~\ref{fig:map1}a.
Their image encompassed 71 of the known members in the current census 
of Chamaeleon~I (\S~\ref{sec:class}), 38 of which were detected. 
One of the X-ray sources from \citet{stel04}, HD~97240 (KG~2001-78), 
was treated as a possible member of Chamaeleon~I in that study. 
However, it is probably a foreground star rather than a cluster member
based on its proper motion and parallax \citep{whi87,per97}.

\begin{figure}
\plotone{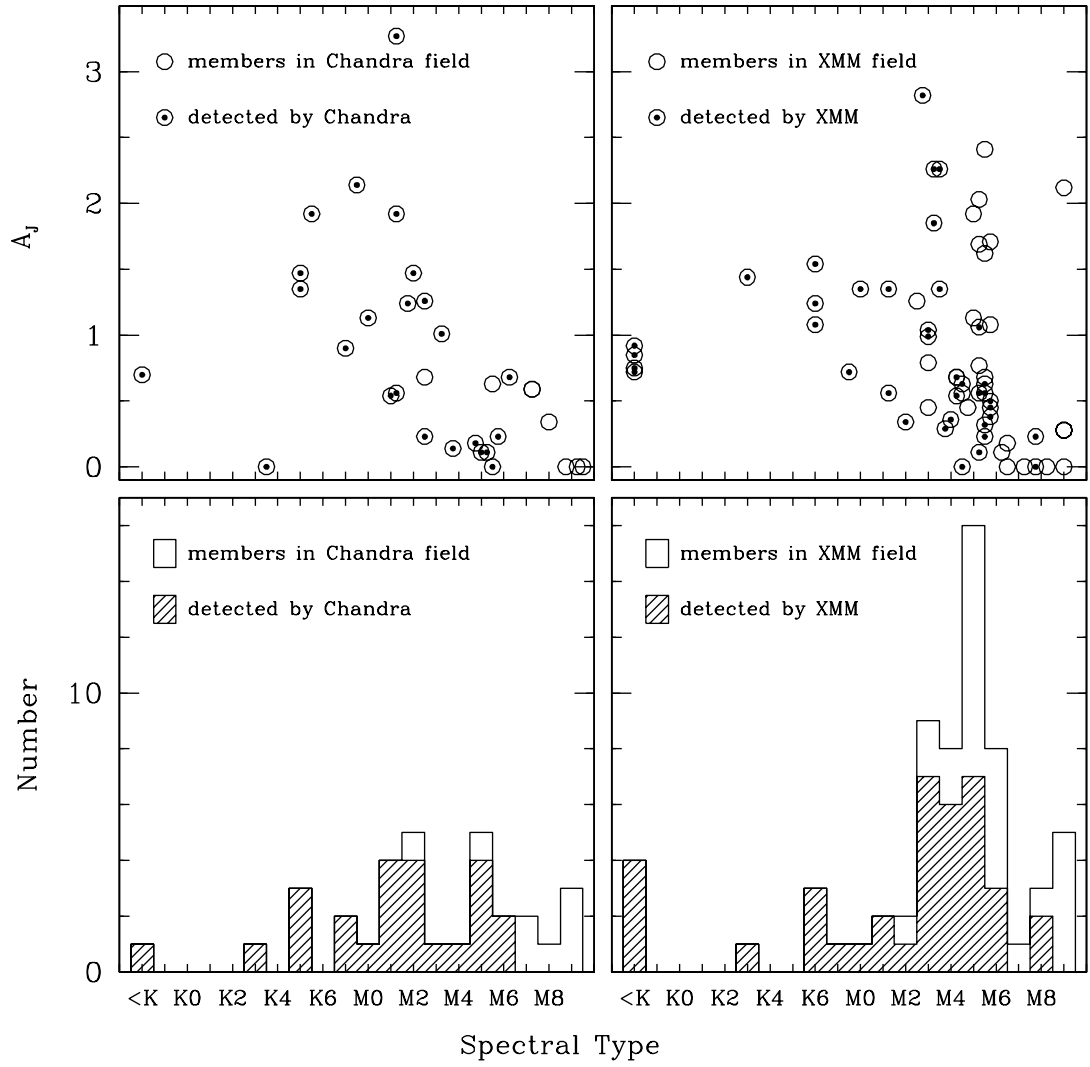}
\caption{
Extinctions ({\it top}) and spectral types ({\it bottom}) for the known members 
of Chamaeleon~I ({\it circles and histograms}) within the {\it Chandra} 
({\it left}) and {\it XMM} ({\it right}) fields shown in 
Figure~\ref{fig:map1}a. The members detected by these X-ray telescopes are 
indicated ({\it circled points and shaded histograms}).
}
\label{fig:histox}
\end{figure}

To investigate the completeness of the X-ray observations
by \citet{fl04} and \citet{stel04}, I show in Figure~\ref{fig:histox}
the distributions of spectral types for all known members of Chamaeleon~I
within each survey field and for the members detected in X-rays. 
Five and three of the known members in the {\it XMM-Newton} and {\it Chandra}
fields, respectively, are not included in the histograms because 
accurate spectral types are not available for them. 
According to Figure~\ref{fig:histox}, the {\it XMM-Newton} data are
complete for spectral types earlier than M2 ($M\gtrsim0.5$~$M_\odot$)
and the level of completeness steadily decreases with later types. 
In comparison, the completeness of the {\it Chandra} image from \citet{fl04} 
changes more abruptly and remains close to 100\% down to $\sim$M6.
To examine the origin of the differences in completeness limits 
between these two X-ray studies, I have included in Figure~\ref{fig:histox}
my estimates of $A_J$ as function of spectral type. 
At M5-M6, the members that were not detected by {\it XMM-Newton} have 
systematically higher extinctions than the detected members, indicating
that extinction prevented their detection. Most of the nondetections at earlier
types were too close to brighter stars to be resolved by
{\it XMM-Newton}, including T39B, T30, and 2MASS~11072022-7738111.
Compared to the {\it XMM-Newton} field, the extinction from the cloud
decreases more rapidly from the center to the edge of the {\it Chandra} field,
as shown in Figure~\ref{fig:map1}. Because the low-mass stars are distributed
more widely than the high-mass stars in the northern subcluster of Chamaeleon~I 
(\S~\ref{sec:spatial})\footnote{For Chamaeleon~I members within the 
{\it Chandra} image, the average distances from the center of the
image are 3.5 and $7\arcmin$ for spectral types of $<$M3 and M3-M6,
respectively.}, they are subjected to lower extinction in the {\it Chandra} 
field than in the {\it XMM-Newton} field. As a result, {\it Chandra}
was able to detect a higher fraction of M5-M6 stars than {\it XMM-Newton}. 

\citet{fl04} measured an IMF from the sample of members detected in their
{\it Chandra} data. Their IMF exhibited a significant deficiency of
low-mass stars (0.1-0.3~$M_\odot$) relative to the Orion Nebula Cluster 
\citep{hil97,hc00,mue02}. The same deficiency is present when the 
{\it Chandra} field is compared to the {\it XMM-Newton} field in the southern 
subcluster and the larger $1.5\arcdeg\times0.35\arcdeg$ area in 
Chamaeleon~I, as shown in Figures~\ref{fig:histo} and \ref{fig:histox}.
The membership lists from \citet{fl04} and this work agree almost perfectly
down to M6, or 0.1~$M_\odot$, so this effect is not a result of incompleteness
in the {\it Chandra} data. Indeed, my census is complete to much lower masses
for most of the {\it Chandra} field, and it still exhibits the deficit of 
low-mass stars (Figure~\ref{fig:histox}). 
Instead of incompleteness, the anomalous IMF in the {\it Chandra} field
is a reflection of the segregation of low-mass stars to larger radii 
surrounding the northern subcluster that was described in \S~\ref{sec:spatial}.

\subsection{\citet{lm04}}

\begin{figure}
\plotone{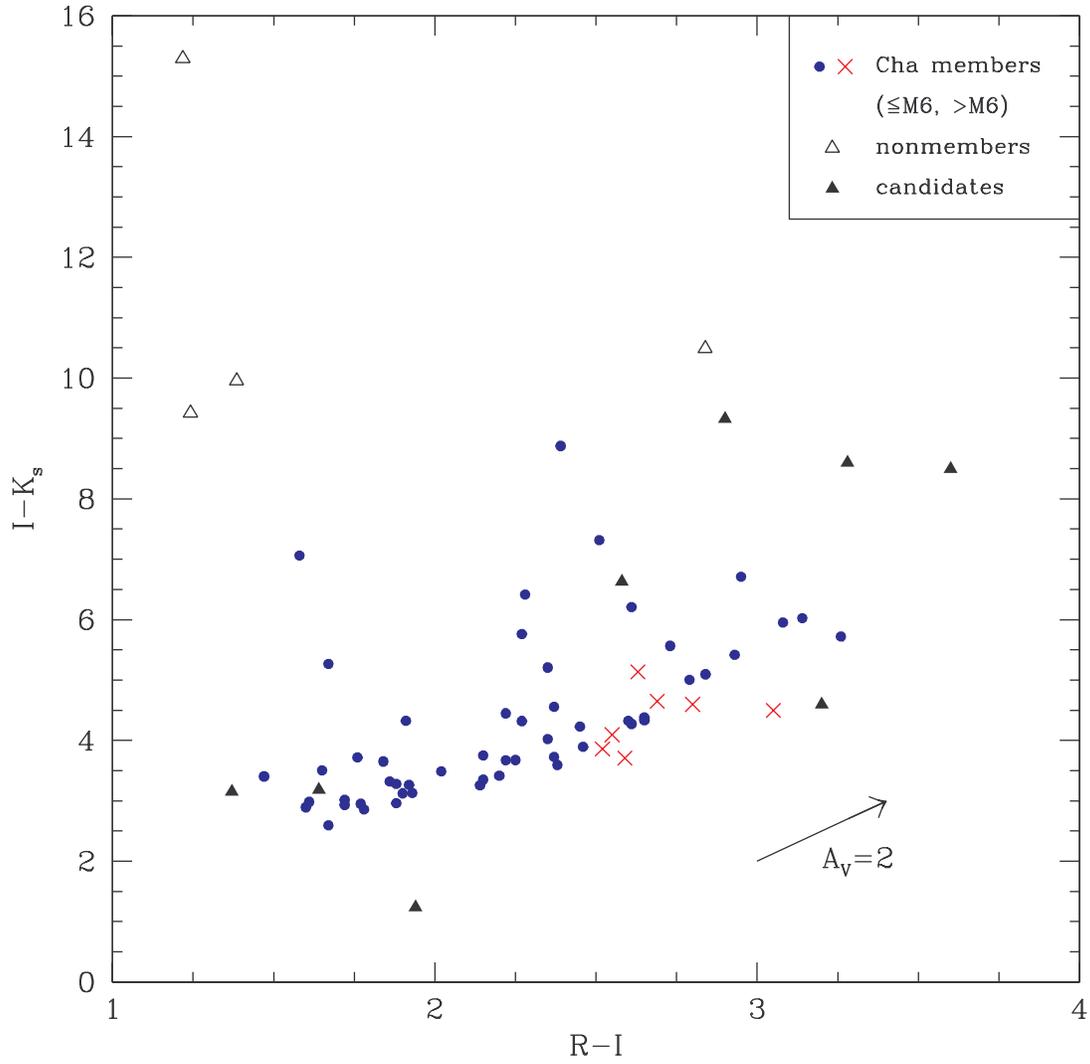}
\caption{
Color-color diagram for the candidate members of Chamaeleon~I identified
by \citet{lm04}. Most of these stars have been spectroscopically classified
as members ({\it filled circles and crosses}) or nonmembers ({\it triangles})
in this work and in previous studies.
Only one of the remaining unclassified candidates exhibits colors that are
indicative of a late-type ($>$M6) member. 
}
\label{fig:riik}
\end{figure}

\citet{lm04} obtained images in $R$, $I$, H$\alpha$, and two narrowband 
optical filters toward 1.2~deg$^2$ in Chamaeleon~I. 
With these data, they identified 69 candidate members of the cluster. 
They selected three additional stars as candidates based on X-ray emission
and IR excess emission.
Among these 72 candidates, 50 objects were spectroscopically classified as 
members by \citet{luh04cha} and \citet{com04} and 10 objects have been 
classified as members in this work. 
Four candidates are field stars based on my spectroscopy (Table~\ref{tab:non}).
The remaining eight candidates lack accurate spectral classifications,
consisting of sources 424, 427, 428, 441, 607, 618, 619, and 749 from 
\citet{lm04}. 
The first five objects were classified as late M or early L by \citet{lm04} 
through their optical colors, while the latter three candidates exhibited 
bluer colors indicative of $<$M4.
These candidates were originally identified by \citet{lm04} through 
a diagram of $I$ versus $R-I$; their candidacy can be further refined by 
plotting them in a diagram of $I-K_s$ versus $R-I$ \citep{luh01}. 
In Figure~\ref{fig:riik}, I show this diagram for the
72 candidates from \citet{lm04}, which are divided into members
at $\leq$M6 and $>$M6, nonmembers, and the eight candidates that lack
spectral classifications. In the latter group, the five objects with
the reddest colors are the ones classified as late M or L by \citet{lm04}.
Only source 424 exhibits colors that are consistent with those of the 
confirmed late-type members. The other four candidates are too red in
$I-K_s$ for their $R-I$, which suggests that they are highly reddened
early-type objects rather than late-type sources. 
This conclusion is supported by the near-IR spectra of sources 428 and 607
from \citet{gm03}, which lack the strong steam absorption that 
is expected for late M and L types.

\citet{lm04} estimated spectral types for their 72 candidates from optical
colors. The classifications of 29 candidates were later than M6. 
In comparison, 15 and seven of these 29 objects have been spectroscopically
classified as members with types of $\leq$M6 and $>$M6, respectively. 
Three late-type candidates are field stars according to my spectroscopy 
and three candidates are likely to have earlier types based on their 
positions in Figure~\ref{fig:riik} and the absence of steam absorption in
the spectra from \citet{gm03}, while source 424 remains a viable late-type
candidate. Thus, no more than eight of the 29 objects that were classified
as later than M6 by \citet{lm04} actually have spectral types in this range. 
It is likely that the types from \citet{lm04} were systematically
too late because of extinction.
When estimating spectral types from the colors of their candidates,
\citet{lm04} did not correct for extinction, which was unknown without
spectroscopic data. 
Because some of their candidates were field stars and their spectral types
exhibited systematic errors, the validity of the IMF derived from that
sample of candidates by \citet{lm04} is questionable.

\subsection{\citet{com04}}

\citet{com04} performed an H$\alpha$ survey across all of Chamaeleon~I 
and obtained spectroscopy of candidate members appearing in those images.
They discovered 18 new members of the cluster, seven and eight of which were 
independently found by \citet{luh04cha} and this study. 
The spectral types from \citet{com99,com00,com04} and \citet{nc99}
are systematically later than those from \citet{luh04cha} and this work by
an average of 0.5~subclass.

\LongTables

\begin{deluxetable}{llllllllllll}
\tabletypesize{\scriptsize}
\tablewidth{0pt}
\tablecaption{Photometry for Candidate Members in Spectroscopic Sample\label{tab:phot}}
\tablehead{
\colhead{Name} &
\colhead{$R$} &
\colhead{$I$} &
\colhead{$i$\tablenotemark{a}} &
\colhead{$m_{775}$} &
\colhead{$m_{850}$} &
\colhead{$J$\tablenotemark{b}} &
\colhead{$H$\tablenotemark{b}} &
\colhead{$K_s$\tablenotemark{b}} &
\colhead{$J$\tablenotemark{c}} &
\colhead{$H$\tablenotemark{c}} &
\colhead{$K_s$\tablenotemark{c}} 
}
\startdata
2M J10314767-7917180 &     \nodata &     \nodata &  15.35 &     \nodata &     \nodata &  13.41 &  12.84 &  12.51 &     \nodata &     \nodata &     \nodata \\
2M J10435748-7633023 &     \nodata &     \nodata &  14.73 &     \nodata &     \nodata &  12.96 &  12.39 &  12.07 &     \nodata &     \nodata &     \nodata \\
2M J10452780-7715335 &     \nodata &     \nodata &  11.44 &     \nodata &     \nodata &   8.65 &   7.65 &   7.17 &     \nodata &     \nodata &     \nodata \\
2M J10523694-7440287 &     \nodata &     \nodata &  13.60 &     \nodata &     \nodata &  11.45 &  10.71 &  10.44 &     \nodata &     \nodata &     \nodata \\
2M J10555824-7418347 &     \nodata &     \nodata &  16.39 &     \nodata &     \nodata &  13.92 &  13.34 &  12.93 &     \nodata &     \nodata &     \nodata \\
2M J10561638-7630530 &     \nodata &     \nodata &  14.92 &     \nodata &     \nodata &  12.54 &  11.94 &  11.52 &     \nodata &     \nodata &     \nodata \\
2M J10580597-7711501 &     \nodata &     \nodata &  15.89 &     \nodata &     \nodata &  13.40 &  12.69 &  12.27 &     \nodata &     \nodata &     \nodata \\
2M J10593228-7753268 &     \nodata &     \nodata &  14.96 &     \nodata &     \nodata &  13.19 &  12.61 &  12.30 &     \nodata &     \nodata &     \nodata \\
\enddata
\tablenotetext{a}{Second DENIS Release.}
\tablenotetext{b}{2MASS Point Source Catalog.}
\tablenotetext{c}{Measured with ISPI in this work.}
\end{deluxetable}

\clearpage
\begin{deluxetable}{clllll}
\tablewidth{0pt}
\tablecaption{Spectroscopy Observing Log\label{tab:log}}
\tablehead{
\colhead{} &
\colhead{} &
\colhead{} &
\colhead{} &
\colhead{$\lambda$} &
\colhead{} \\
\colhead{Night} &
\colhead{Date} & 
\colhead{Telescope + Instrument} &
\colhead{Disperser} &
\colhead{($\mu$m)} &
\colhead{$\lambda/\Delta \lambda$}
}
\startdata
1 & 2004 Jan 2 & Magellan~II + MIKE & \nodata & 0.5-1 & 20000 \\
2 & 2004 Jan 7 & Magellan~I + IMACS & 200 grism & 0.5-1 & 900 \\
3 & 2004 Jan 8 & Magellan~I + IMACS & 200 grism & 0.5-1 & 900 \\
4 & 2004 Jan 9 & Magellan~I + IMACS & 200 grism & 0.5-1 & 900 \\
5 & 2004 Jan 10 & Magellan~I + IMACS & 200 grism & 0.5-1 & 900 \\
6 & 2004 Jan 11 & Magellan~I + IMACS & 200 grism & 0.5-1 & 900 \\
7 & 2004 Mar 31 & Magellan~II + LDSS-2 & med red grism & 0.48-0.92 & 600 \\
8 & 2004 Apr 26 & Magellan~I + IMACS & 200 grism & 0.5-1 & 900 \\
9 & 2004 Apr 27 & Magellan~I + IMACS & 200 grism & 0.5-1 & 900 \\
10 & 2005 Jan 2 & Magellan~I + IMACS & 200 grism & 0.5-1 & 1500 \\
11 & 2005 Jan 3 & Magellan~I + IMACS & 200 grism & 0.5-1 & 900 \\
12 & 2005 Jan 4 & Magellan~I + IMACS & 200 grism & 0.5-1 & 900 \\
13 & 2005 Jan 5 & Magellan~I + IMACS & 200 grism & 0.5-1 & 900 \\
14 & 2005 Jan 6 & Magellan~I + IMACS & 200 grism & 0.5-1 & 900 \\
15 & 2005 Mar 23 & Gemini + GNIRS & 31.7 cross-dispersed & 1-2.5 & 2000 \\
16 & 2005 Mar 24 & Gemini + GNIRS & 31.7 cross-dispersed & 1-2.5 & 2000 \\
17 & 2005 Mar 25 & Gemini + GNIRS & 31.7 cross-dispersed & 1-2.5 & 2000 \\
\enddata
\end{deluxetable}

\begin{deluxetable}{llllll}
\tabletypesize{\scriptsize}
\tablewidth{0pt}
\tablecaption{New Members of Chamaeleon I\label{tab:new}}
\tablehead{
\colhead{} &
\colhead{} &
\colhead{Spectral} &
\colhead{Membership} &
\colhead{Basis of} &
\colhead{} \\
\colhead{2MASS\tablenotemark{a}} &
\colhead{Name\tablenotemark{b}} &
\colhead{Type\tablenotemark{c}} &
\colhead{Evidence\tablenotemark{d}} &
\colhead{Selection\tablenotemark{e}} &
\colhead{Night}} 
\startdata
J10523694-7440287 &         \nodata &      M4.75 &        NaK &      IK2 &        7   \\
J10561638-7630530 &  ESO H$\alpha$ 553 &       M5.6 &     NaK,e &      IK1 &        4   \\
J10580597-7711501 &         \nodata &      M5.25 &        NaK &      IK1 &        4   \\
J11011370-7722387 &         \nodata &      M5.25 &        NaK &      IK1 &        7   \\
J11021927-7536576 &         \nodata &       M4.5 &        NaK &      IK2 &        7   \\
J11022610-7502407 &         \nodata &      M4.75 &        NaK &      IK2 &        7   \\
J11024183-7724245 & [LES2004] 423 &         M5 &     NaK,$A_V$ &      IK1 &        7   \\
J11050752-7812063 &         \nodata &      M4.75 &        NaK &      IK1 &        7   \\
\enddata
\tablenotetext{a}{2MASS Point Source Catalog.}
\tablenotetext{b}{Sources that are not in the 2MASS Point Source Catalog 
have been assigned coordinate-based identifications.}
\tablenotetext{c}{Uncertainties are $\pm0.25$ subclass unless noted otherwise.}
\tablenotetext{d}{Membership in Chamaeleon~I is indicated by $A_V\gtrsim1$ and
a position above the main sequence for the distance of Chamaeleon~I (``$A_V$"),
strong emission lines (``e"), Na~I and K~I strengths intermediate
between those of dwarfs and giants (``NaK"), IR excess emission (``ex"),
or the shape of the gravity-sensitive steam bands (``H$_2$O").}
\tablenotetext{e}{Sources were selected for spectroscopy because 
they appeared as candidate members in
Figure~\ref{fig:ik1} (``IK1"), Figure~\ref{fig:ik2} (``IK2"), 
Figure~\ref{fig:ri} (``RI"), or Figure~\ref{fig:iz} (``IZ") or 
were identified as candidate members by \citet{lm04} (``LES2004").}
\tablenotetext{f}{Possible wide companion to T26 \citep{rz93,ghe97}.}
\end{deluxetable}

\begin{deluxetable}{lllll}
\tabletypesize{\scriptsize}
\tablewidth{0pt}
\tablecaption{Nonmembers\label{tab:non}}
\tablehead{
\colhead{} &
\colhead{} &
\colhead{Spectral} &
\colhead{Basis of} &
\colhead{} \\
\colhead{2MASS\tablenotemark{a}} &
\colhead{Name\tablenotemark{b}} &
\colhead{Type\tablenotemark{c}} &
\colhead{Selection\tablenotemark{d}} &
\colhead{Night}} 
\startdata
J10314767-7917180 &         \nodata &     M5.25V &      IK2 &        7   \\
J10435748-7633023 &         \nodata &     M4.75V\tablenotemark{e} &      IK2 &        7   \\
J10452780-7715335 &         \nodata &      M6III &      IK1 &        9   \\
J10555824-7418347 &         \nodata &      M7.5V\tablenotemark{e} &      IK2 &        7   \\
J10593228-7753268 &         \nodata &     M4.75V &      IK1 &        7   \\
J11001473-7757103 &         \nodata &        $<$K0 &      IK1 &        7   \\
J11014949-7412268 &         \nodata &          giant &      IK2 &        7   \\
J11035902-7743349 &     ISO 35 &        $<$K0 &       RI &       14   \\
\enddata
\tablenotetext{a}{2MASS Point Source Catalog.}
\tablenotetext{b}{Sources that are not in the 2MASS Point Source Catalog 
have been assigned coordinate-based identifications.}
\tablenotetext{c}{Uncertainties are $\pm0.25$ subclass unless noted otherwise.}
\tablenotetext{d}{Sources were selected for spectroscopy because 
their previous classifications were uncertain (``unc"), they were previously
identified candidate companions (``comp"), they were candidate members based on 
Figure~\ref{fig:ik1} (``IK1"), Figure~\ref{fig:ik2} (``IK2"), 
Figure~\ref{fig:ri} (``RI"), Figure~\ref{fig:iz} (``IZ"), 2MASS photometry 
(``2M"), {\it Spitzer} photometry (``IRAC"), or X-ray emission (``X-ray"), or 
they were identified as candidate members by \citet{lm04} (``LES2004").}
\tablenotetext{e}{Spectral features suggest a surface
gravity intermediate between those of Chamaeleon~I members and field dwarfs.}
\tablenotetext{f}{Classified as a background giant by \citet{bz97}.}
\tablenotetext{g}{Classified as an M-type member of Chamaeleon~I by \citet{gm03}.}
\tablenotetext{h}{Classified as M5.5V by \citet{com04}.}
\end{deluxetable}

\begin{deluxetable}{llllll}
\tabletypesize{\scriptsize}
\tablewidth{0pt}
\tablecaption{Previously Known Members of Chamaeleon I with New Spectroscopy\label{tab:old}}
\tablehead{
\colhead{} &
\colhead{} &
\colhead{Spectral} &
\colhead{} &
\colhead{Membership} &
\colhead{} \\
\colhead{2MASS\tablenotemark{a}} &
\colhead{Name} &
\colhead{Type\tablenotemark{b}} &
\colhead{Ref} &
\colhead{Evidence\tablenotemark{c}} &
\colhead{Night}} 
\startdata
J10550964-7730540 &    ESO H$\alpha$ 552 &    M5,M4.5 &    1,2 &      NaK &       10   \\
J10555973-7724399 &        T3A & M1,M0.5,M0.5,K8-M0.5 & 3,4,5,2 & e,ex,rv,$A_V$ &        4   \\
              \nodata &        T3B &       M3.5 &        2 &      NaK &        4   \\
J11013205-7718249\tablenotemark{d} &    ESO H$\alpha$ 554 &    M8.5,M8 &    1,2 &      NaK &        5   \\
J11062554-7633418\tablenotemark{d} &    ESO H$\alpha$ 559 &   M6,M5.25 &    1,2 & NaK,e,$A_V$,ex &        5   \\
J11065906-7718535 &        T23 & M1.5,M4.25 &   6,2 & e,ex,NaK &        4   \\
J11070925-7718471 &      ISO 91 & M2.5,M2-M4 &   7,2 &       ex &        3   \\
J11072825-7652118 &        T27 & K0:,M1,M1.5,M1.5+M3,M3 & 3,4,5,8,2 & e,ex,rv,Li &        4   \\
\enddata
\tablenotetext{a}{2MASS Point Source Catalog.}
\tablenotetext{b}{Uncertainties for classifications from this work 
are $\pm0.25$ subclass unless noted otherwise.}
\tablenotetext{c}{Membership in Chamaeleon~I is indicated by $A_V\gtrsim1$ and
a position above the main sequence for the distance of Chamaeleon~I (``$A_V$"),
strong emission lines (``e"), Na~I and K~I strengths intermediate
between those of dwarfs and giants (``NaK"), IR excess emission (``ex"),
the shape of the gravity-sensitive steam bands (``H$_2$O"), 
strong Li absorption (``Li"), or a radial velocity that is similar 
to that of the known members of Chamaeleon~I (``rv").}
\tablenotetext{d}{Selected for spectroscopy because it appeared as a candidate 
member in Figure~\ref{fig:ik1}.}
\tablenotetext{e}{Separation of $1.1\arcsec$ and position angle of $19\arcdeg$
relative to T39A.}
\tablerefs{
(1) \citet{com04};
(2) this work;
(3) \citet{app79};
(4) \citet{ryd80};
(5) \citet{app83};
(6) \citet{gs92};
(7) \citet{gm03};
(8) \citet{bz97};
(9) \citet{com99};
(10) \citet{fk89};
(11) \citet{hlf94};
(12) \citet{alc95};
(13) \citet{cov97};
(14) \citet{saf03};
(15) \citet{luh04cha};
(16) \citet{luh04ots};
(17) \citet{har93}.
}
\end{deluxetable}

\begin{deluxetable}{lllllll}
\tabletypesize{\scriptsize}
\tablewidth{0pt}
\tablecaption{Derived Properties for Members of Chamaeleon I\label{tab:hr}}
\tablehead{
\colhead{} &
\colhead{} &
\colhead{Adopted} &
\colhead{$T_{\rm eff}$\tablenotemark{b}} &
\colhead{} &
\colhead{$L_{\rm bol}$} &
\colhead{} \\
\colhead{2MASS\tablenotemark{a}} &
\colhead{Name} &
\colhead{Spectral Type} &
\colhead{(K)} &
\colhead{$A_J$} &
\colhead{($L_\odot$)} &
\colhead{IMF\tablenotemark{c}}}
\startdata
J10463795-7736035 &    HD 93828 &         F0 &  7200 &  0.00 &     8.9 &  \nodata \\   
J10523694-7440287 &         \nodata &      M4.75 &  3161 &  0.18 &    0.10 &  \nodata \\   
J10550964-7730540 &    ESO H$\alpha$ 552 &       M4.5 &  3198 &  0.34 &   0.083 &  \nodata \\   
              \nodata &        T3B &       M3.5 &  3342 &  0.00 &   0.040 &  \nodata \\   
J10555973-7724399 &        T3A &         M0 &  3850 &  0.79 &    0.42 &  \nodata \\   
J10561638-7630530 &   ESO H$\alpha$ 553 &       M5.6 &  3044 &  0.00 &   0.031 &  \nodata \\   
J10563044-7711393 &         T4 &       M0.5 &  3778 &  0.23 &    0.57 &  \nodata \\   
J10574219-7659356 &         T5 &      M3.25 &  3379 &  0.34 &    0.33 &  \nodata \\
\enddata
\tablenotetext{a}{2MASS Point Source Catalog.}
\tablenotetext{b}{Temperature scale from \citet{luh03ic}.}
\tablenotetext{c}{Indicates whether each object is included in the IMFs in
Figure~\ref{fig:imf} for the $1.5\arcdeg\times0.35\arcdeg$ (``1") and
$0.22\arcdeg\times0.28\arcdeg$ (``2") fields in Figure~\ref{fig:map1}.}
\end{deluxetable}

\clearpage

\clearpage

\clearpage

\end{document}